\documentclass[12pt,titlepage]{article}
\pdfoutput=1
\usepackage{jheppub}
\usepackage{color}
\usepackage{graphicx}
\numberwithin{equation}{section}
\DeclareSymbolFont{AMSb}{U}{msb}{m}{n}
\DeclareMathSymbol{\IN}{\mathbin}{AMSb}{"4E}
\DeclareMathSymbol{\IZ}{\mathbin}{AMSb}{"5A}
\DeclareMathSymbol{\IR}{\mathbin}{AMSb}{"52}
\DeclareMathSymbol{\Q}{\mathbin}{AMSb}{"51}
\DeclareMathSymbol{\II}{\mathbin}{AMSb}{"49}
\DeclareMathSymbol{\IC}{\mathbin}{AMSb}{"43}
\DeclareMathSymbol{\IP}{\mathbin}{AMSb}{"50}
\DeclareMathSymbol{\IH}{\mathbin}{AMSb}{"48}
\DeclareMathSymbol\IA{\mathalpha}{AMSb}{"41}
\DeclareMathSymbol\IS{\mathalpha}{AMSb}{"53}

\def\matt[#1,#2,#3,#4]{\left(%
\begin{array}{cc} #1 & #2 \\ #3 & #4 \end{array} \right)}

\def\v2#1{\vv2[#1]}
\def\vv2[#1,#2]{\left(%
{#1 \atop #2}\right)}

\begin{document}

\preprint{DIAS-STP-12-10\\}

\title{{\boldmath Flavoured Large N Gauge Theory on a Compact Space with an External Magnetic Field}}

\author[a]{Veselin G. Filev,}
\author[a]{Matthias Ihl}
\affiliation[a]{  School of Theoretical Physics,\\ 
       Dublin Institute for Advanced Studies, \\
       10 Burlington Road, 
       Dublin 4, Ireland.}
\emailAdd{vfilev@stp.dias.ie}
\emailAdd{msihl@stp.dias.ie}
\abstract{
The phase structure of flavoured ${\cal N}=2$ SYM on a three sphere in an external magnetic field is studied. 
The pairing effect of the magnetic field competes with the dissociating effect of the Casimir energy, leading to an interesting phase structure of 
confined and deconfined phases separated by a critical curve of a first order quantum phase transition. 
At vanishing magnetic field the phase transition is of a third order. For sufficiently strong magnetic field, 
the only stable phase is the confined phase and magnetic catalysis of chiral symmetry breaking is realized. 
The meson spectra of the theory exhibit Zeeman splitting and level crossing and feature a finite jump at the phase transition between the confined and 
deconfined phases. At strong magnetic field the ground state has a massless mode corresponding to the Goldstone boson associated with the spontaneously 
broken $U(1)$ $R$-symmetry analogous to the $\eta'$ meson in QCD.}

\maketitle
\newpage
\section{Introduction}
The influence of magnetic fields on flavoured gauge theories has been extensively studied in the literature. 
In the pioneering works of refs.\cite{Gusynin:1994re}-\cite{Klimenko:1992ch}, it has been shown that magnetic fields act as a strong catalyst of mass generation and chiral symmetry breaking 
leading to the formation of a fermionic condensate, even in the slightest attractive potential. 
The essence of this effect is the dimensional reduction, from D to D--2 dimensions, in the dynamics of fermion pairing. 
The effect has been shown to be model independent and therefore has a universal nature.
Given such universal nature, it is natural to study this effect in holographic gauge theories, 
where one can rely on the powerful techniques of the AdS/CFT correspondence at strong coupling. 

In its original formulation \cite{Maldacena:1997re}, the correspondence relates perturbative string theory on an AdS$_5\times S^5$ background to a four dimensional ${\cal N}=4$ SYM theory.
An important extension of the correspondence \cite{Karch:2002sh}, making it relevant to the description of flavoured Yang-Mills theories, 
was the inclusion of fundamental matter via the introduction of flavour branes in the so-called probe limit, where the number of different flavours is much less 
than the number of colors, $N_f \ll N_c$. This corresponds to the quenched approximation on the gauge theory side and the probe approximation on the 
supergravity side of the correspondence. 

The holographic approach to magnetic catalysis was initiated in ref.~\cite{Filev:2007gb}, where the holographic gauge theory dual to the D3/D7 
intersection has been analyzed. Further relevant studies have been performed in refs. \cite{Filev:2011mt}-\cite{Chernodub:2010qx}. 
Holographic studies of backreacted flavours in external magnetic fields have been performed in refs. \cite{Filev:2011mt}-\cite{Ammon:2012qs}.

In the present work, we are interested in studying the effect of an external magnetic field on a flavoured gauge theory defined on a compact space. 
A canonical example of holographic gauge theory on compact spaces is ${\cal N}=4$ SYM theory on a three-sphere which is dual to string theory defined on 
AdS$_5\times S^5$ in global coordinates. 
However, the second Betti number of the three-sphere is zero, suggesting that a non-zero magnetic field on the three-sphere requires a non-zero electric current. The full treatment of this problem in a top-down approach requires a complex ansatz for the magnetic field\footnote{See ref.~\cite{Chunlen:2014zpa} for studies in this direction.}. In this paper we take a different approach by considering a semi-bottom up model, in which the electric current required to support the magnetic field is introduced by an external seven form Ramond-Ramond flux, whose backreaction we ignore.

The addition of fundamental flavours is achieved in the same way as in the flat case, namely by introducing probe D--branes. 
The confinement/deconfinement phase transition of flavours on $S^4$ has been studied in refs.~\cite{Karch:2006bv,Karch:2009ph,Erdmenger:2010zm}, 
and the effect of $R$-charge and isospin chemical potentials was addressed in refs.~\cite{PremKumar:2011ag,Chunlen:2012zy}.

The finite volume gives and extra energy scale associated to the Casimir energy of the theory. 
Despite the fact that our study is at zero temperature, the expected effect of the Casimir free energy is to favour the dissociation of meson-like bound states,
triggering a confinement/deconfinement phase transition \cite{Karch:2009ph}. On the other hand, magnetic fields favour the formation of bound states. 
Therefore, one would expect to find an interesting phase structure of confined and deconfined phases separated by a critical curve across 
which a quantum confinement/deconfinement phase transition takes place. 
Note that this phase transition does not have an analogue in the theory on a flat space. 
However, for sufficiently strong magnetic fields the effect of the Casimir energy will be subdominant and the theory should be qualitatively similar to the flat case. 
Thus we expect that for strong magnetic fields the phase transition seizes to exist, and the theory is in a confined phase where the vacuum at zero bare mass 
spontaneously breaks a global flavour symmetry.

The paper is organised as follows: 
In section 2, we review the holographic setup describing ${\cal N}=2$ SYM on an $S^3$, studied in refs.~\cite {Karch:2006bv, Karch:2009ph, Erdmenger:2010zm}. 
We define a set of coordinates convenient for the study of probe branes and their classification. 
Furthermore, we comment on the AdS/ CFT dictionary and the holographic renormalization of the probe branes, introduce the scaling exponents 
characterizing the self-similar behaviour of the theory near the topology changing transition of the D7--brane embeddings and 
review the calculation of the order of the phase transition \cite{Karch:2009ph}.

In section 3.1 and 3.2, an external magnetic field along two of the $S^3$ directions is introduced. 
We construct the D7--brane embeddings and study the self-similar regime of the theory, extracting complex scaling exponents. 
The analysis of scaling exponents is supplemented with numerical results to show that in an external magnetic field 
the phase transition is of first order. Next, we study the dependence of the fundamental condensate on the bare mass parameter for various magnetic fields. 
The effect of the magnetic field is to decrease the critical mass at which the confinement/deconfinement phase transition takes place. 
For sufficiently large magnetic field the critical mass vanishes and the transition happens at zero bare mass. 
Beyond this point there is no phase transition at all and at vanishing bare mass the stable vacuum features a non-vanishing negative condensate that spontaneously breaks 
the axial U(1) R-symmetry. To illustrate this, we construct a phase diagram of the theory summarizing this behavior. 

In section 3.3, the meson spectra of the theory are analysed. The effect of the magnetic field is to couple the vector and scalar modes.
We show that the self-similar behaviour of the theory at finite magnetic field has a tachyonic instability. Across the phase transition, 
the meson spectrum displays a finite jump between the confined and deconfined phases. Moreover, we demonstrate that the globally symmetric vacuum develops 
a tachyonic instability for strong magnetic fields. However, it continues to be metastable for a small window of values of the magnetic field, even though the 
stable phase of the theory is the one with spontaneously broken global symmetry and negative condensate. 
The spectrum of the mixed modes exhibits Zeeman splitting at large bare masses, which leads to level crossing in the confined phase of the theory.
For strong magnetic fields, the phase transition seizes to exist and the theory is in the confined phase. 
The ground state of the spectrum possesses a massless mode corresponding to the Goldstone boson associated with the spontaneous breaking of the global 
$U(1)$ $R$--symmetry of the theory, analogous to the $\eta$' meson in QCD. 
Furthermore, by studying the dependence of the mass of the meson on the bare quark mass near the origin, we find a characteristic $M\propto\sqrt {m}$ behaviour 
reminiscient of the Gell-Mann-Oakes-Renner relation \cite{GellMann:1968rz}, corroborating the existence of a massless Goldstone boson.

\section{Review of the model}

\subsection{Flavours on $S^3$}
In this section we review the holographic setup describing flavoured ${\cal N}=2$ SYM on an $S^3$, studied in refs.~\cite{Karch:2006bv, Karch:2009ph, Erdmenger:2010zm}.
Let us consider the metric of AdS$_5\times S^5$ in global coordinates:
\begin{eqnarray}
ds^2&=&-(1+r^2/R^2)d\tau^2+r^2d\Omega_3^{(1)\,2}+\frac{dr^2}{1+r^2/R^2}+R^2d\Omega_5^2\ , \label{Glob}\\
d\Omega_5^2&=&d\theta_3^2+\cos^2\theta_3 d\Omega_3^{(2)\,2}+\sin^2\theta_3 d\phi_3^2\ . \nonumber
\end{eqnarray}
It is convenient to define a new radial coordinate,
\begin{equation}
u=\frac{1}{2}(r+\sqrt{R^2+r^2})\ .
\end{equation}
Then the metric (\ref{Glob}) becomes
\begin{equation}
ds^2=-\frac{u^2}{R^2}\left(1+\frac{R^2}{4u^2}\right)^2d\tau^2+{u^2}\left(1-\frac{R^2}{4u^2}\right)^2d\Omega_3^{(1)\,2}+\frac{R^2}{u^2}\left(du^2+u^2d\Omega_5^2\right)\ .\label{Globu}
\end{equation}
Note that in this radial coordinate the metric has a conformal $\IR^6$ piece. However, $u$ is bounded from below by $u\ge R/2$ and the transverse $\IR^6$ has a ball of radius $R/2$ at the origin. 

Next, we consider a stack of D7--branes embedded along the AdS$_5$ part of the geometry and wrapping an $\tilde S^3\subset S^5$. The dual field theory is ${\cal N}=4$ SYM on $S^3$ coupled to ${\cal N}=2$ hypermultiplets. Note that, unlike the theory on flat space,
a non-vanishing mass of the hypermultiplets breaks all supersymmetries and the theory preserves ${\cal N}=2$ supersymmetry only for massless flavours. In the gravity setup this can be seen by studying the $\kappa$-symmetry condition for the probe branes. 

\begin{figure}[htbp] 
   \centering
   \includegraphics[width=4.5in]{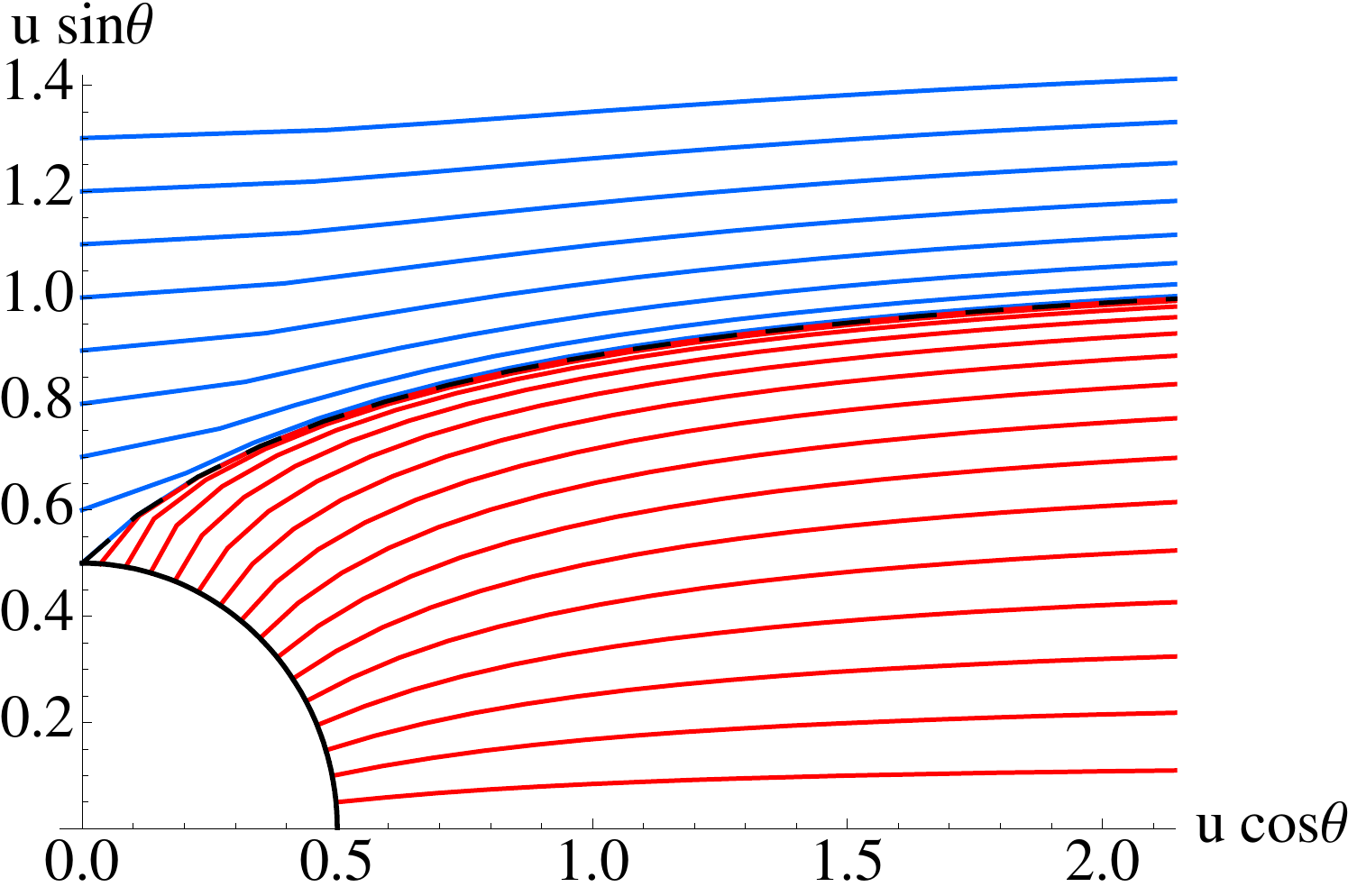} 
   \caption{\small Blue curves correspond to ``Minkowski" embeddings and Red curves correspond to ``ball" embeddings. The black dashed line corresponds to the critical embedding.}
   \label{fig:fig1}
\end{figure}

The shape of the D7--brane embeddings can be determined by extremising the Dirac--Born--Infeld action,
\begin{eqnarray}
S_{\rm{DBI}}=-N_f\mu_{7}\int\limits_{{\cal M}_{8}}d^{8}\xi e^{-\Phi}[-{\rm det}(G_{ab}+B_{ab}+(2\pi\alpha')F_{ab}]^{1/2}\  . \label{DBI}
\end{eqnarray}
The radial part of the corresponding DBI lagrangian is given by,
\begin{equation}
{\cal L} \propto \left(1-\frac{R^4}{16u^4}\right)\left(1-\frac{R^2}{4u^2}\right)^2u^3\cos^3\theta_3\sqrt{1+u^2\theta_3'(u)^2}\ .\label{lagr}
\end{equation}
The resulting D7--brane embeddings fall into two classes \cite{Karch:2006bv}. Embeddings, which reach the origin of the AdS terminate on the shrinking $S^3 \subset$ AdS$_5$. We will call these embeddings ``ball'' embeddings, because they end at $u=R/2$ in terms of the radial coordinate $u$. 
The second class of embeddings terminates above the origin of the AdS and instead wraps a shrinking $\tilde S^3\subset S^5$. We will refer to this class as ``Minkowski'' embeddings  (cf.~figure \ref{fig:fig1}). 
The two classes are separated by a critical embedding which has a conical singularity located on the ball (the origin of AdS). 
The AdS/CFT dictionary relates the asymptotic behaviour of the transverse scalar $\theta_3(u)$ to the {\it vev} and source of the fundamental bilinear, with the source being the bare mass of the hypermultiplet. 
The dictionary has been derived in ref.~\cite{Karch:2005ms} using an appropriate holographic renormalization procedure. For large $u$ one obtains the following expansion,
\begin{equation}
\sin\theta_3=\frac{m}{u}+\frac{c_1}{u^3}-\frac{m}{2u^2}\log\,u+\dots \ .
\end{equation}
The condensate of the theory is then given by,
\begin{equation}
\langle\bar\psi\psi\rangle\propto-2c_1+m\log (m/R)\equiv -2c\ , \label{condcal}
\end{equation}
where we have defined a new parameter $c$ proportional to the condensate. Note also that the bare mass of the hypermultiplet is proportional to $m$: the exact relation is $m_q=m/2\pi\alpha'$. 

\begin{figure}[htbp] 
   \centering
   \includegraphics[width=4.5in]{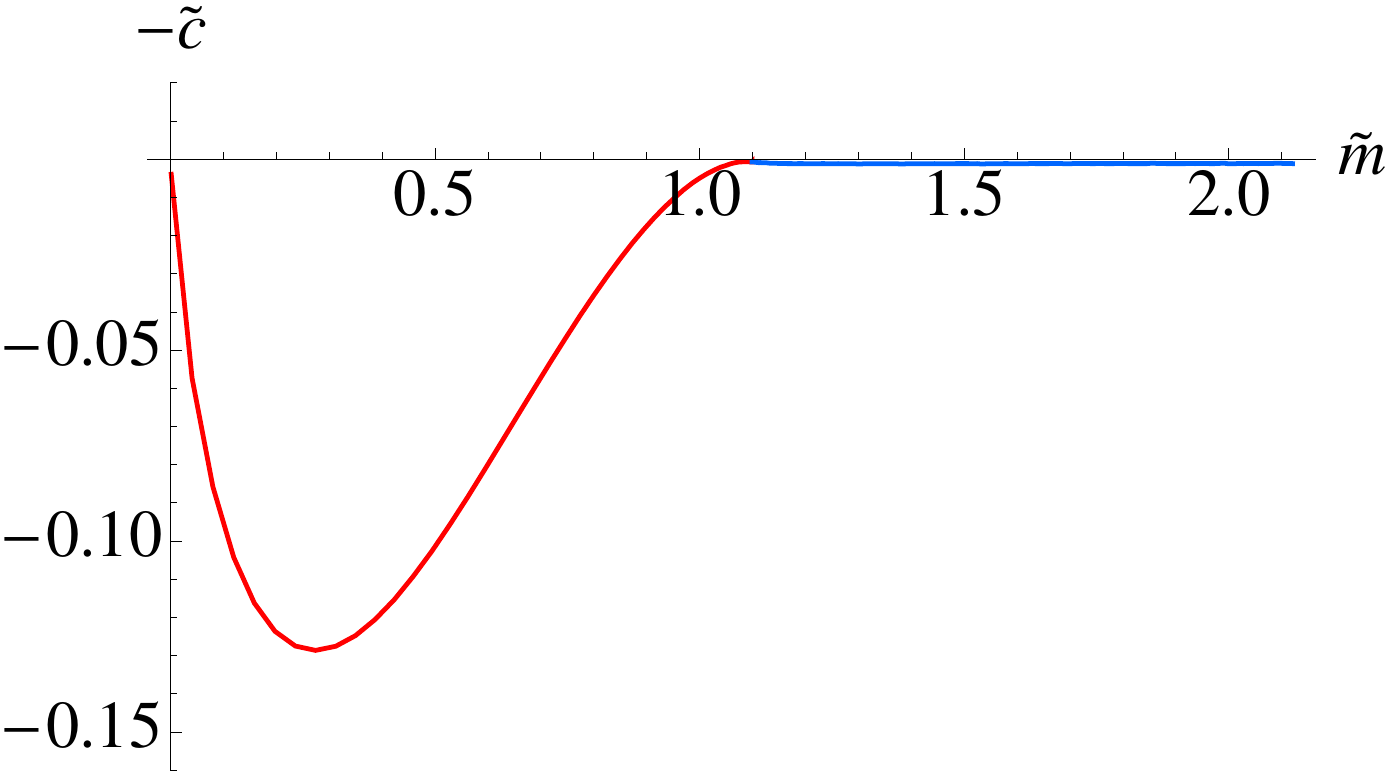} 
   \caption{\small A plot of the condensate $-\tilde c$ versus the bare mass $\tilde m$. The states corresponding to ``ball" embeddings are represented by the red curve and the states corresponding to ``Minkowski'' embeddings are represented by the blue curve.}
   \label{fig:fig2}
\end{figure}

By solving numerically for the D7--brane embeddings and studying the asymptotic behaviour of $\theta_3(u)$, one can construct a plot of the equation of state (i.e., a plot of $c$ versus $m$). We generated this plot in figure \ref{fig:fig2}, 
where we have used dimensionless parameters $\tilde m=m/R$ and $\tilde c=c/R^3$.
Note that there is no apparent multi-valued region near the transition from ``Minkowski" to ``ball" embeddings. In fact, by calculating appropriate critical exponents the authors of ref.~\cite{Karch:2009ph} have shown that it is a third order phase transition. Furthermore, the authors argued that the phase transition corresponds to a quantum confinement/deconfinement phase transition triggered by the Casimir energy of the $S^3$, which destroys the bound states of the quarks for sufficiently large curvature of $S^3$. 
The state for which bound states exist corresponds to values of the parameter $\tilde m$ greater than the critical mass $\tilde m_{\ast}$ (the blue branch of the curve in figure \ref{fig:fig2}). The dissociated phase corresponds to values of the parameter $\tilde m$ smaller than $\tilde{m}_{\ast}$ (the red branch in figure \ref{fig:fig2}). 
The physical meaning of the parameter $\tilde m$ is not obvious from its definition since the scale $R$ represents the radius of $S^3$, but is also related to the t'Hooft coupling $\lambda$ of the theory. The right way of thinking about $\tilde m$ is as follows \cite{Karch:2009ph,Erdmenger:2010zm}: $\tilde m=m/R=m R_3/R^2$, where $R_3$ denotes the radius of $S^3$ and $R^2\propto \sqrt{\lambda}$. 
The exact expression for $\tilde m$ in terms of field theory quantities is
\begin{equation}
\tilde m=\frac {\pi}{\sqrt{2}}\frac{m_q\,R_3}{\sqrt{\lambda}}\ .
\end{equation}
For fixed bare quark mass $m_q$, small values of $\tilde m$ correspond to small radius of $S^3$ and large Casimir energy triggering the dissociation of the bound quarks. Conversely, 
for large $\tilde m$ the radius of $S^3$ is large and the Casimir energy is small relative to the bare quark mass and thus the bound state is stable.

\subsection{Critical exponents and the order of the phase transition}
In this subsection we focus on the vicinity of the critical embedding separating the ``Minkowski'' and ``ball" classes of embeddings. We obtain the corresponding critical exponents and show (along the lines of ref.~\cite{Karch:2009ph}) that the topology changing transition corresponds to a third order phase transition in the dual gauge theory. 
These studies will be applied in section 3 to study the critical exponent of the transition in the presence of external magnetic field.

To begin with, we zoom into the geometry near the tip of the critical embedding (the dashed curve in figure \ref{fig:fig1}). Let us consider the change of variables,
\begin{equation}
u=1/2(1+z);~~y=\pi/2-\theta_3;\label{zooming}
\end{equation}
and expand the metric (\ref{Globu}). To leading order in $z$ and $y$, we obtain
\begin{equation}
ds_{\rm{zoom}}^2=-d\tau^2+z^2d\Omega_3^{(1)\,2}+dz^2+dy^2+y^2d\Omega_3^{(2)\,2}+d\phi_3^2\ ,\label{metricZ}
\end{equation}
which is just a flat metric on $\IR^{1,9}$. The Lagrangian for the D7--brane embedding (\ref{lagr}) becomes
\begin{equation}
{\cal L}_{\rm{zoom}}=z^3y^3\sqrt{1+y'^2}\ .\label{lagrZoom}
\end{equation}
The Lagrangian (\ref{lagrZoom}) is a special case of the general Lagrangian
\begin{equation}
z^{k/2}y^n\sqrt{1+y'^2}\ , \label{lagrgen}
\end{equation}
considered in ref.~\cite{Filev:2008xt}, for $k=6,n=3$. The solutions to the corresponding equation of motion have a scaling property $y(z)\to\frac{1}{\mu}y(\mu z)$ and a critical solution $y_*(z)=\sqrt{2n/k}z$. 
Expanding in the vicinity of the critical solution $y(z)=y_*(z)+\xi(z)$ results in the following equation of motion \cite{Filev:2008xt},
\begin{equation}
z^2\xi''(z)+(n+k/2)(z\xi'(z)+\xi(z))=0\ . \label{EOMZoom}
\end{equation}
Equation (\ref{EOMZoom}) has a general solution
\begin{equation}
\xi(z)={\rm const}_1z^{\delta_1}+{\rm const}_2 z^{\delta_2}\ ,\label{XI}
\end{equation}
where
\begin{equation}
\delta_{1/2}=1/2\left[-(k/2+n-1)\pm\sqrt{(k/2+n-1)^2-4(k/2+n)}\right]\ .\label{deltapm}
\end{equation}
It is easy to see that for $-2\sqrt{2}+3-n<k/2< 2\sqrt{2}+3-n$ the critical exponents $\delta_{1/2}$ are complex. In this window the solution exhibits a discrete self-similar behavior which seeds multi-valuedness in the equation of state and suggests that the phase transition is of first order (for more details,cf. refs~\cite{Karch:2009ph,Filev:2008xt,Mateos:2007vn,Mateos:2006nu,Frolov:2006tc}). 
In section 3 we will see that this holds true in the presence of a magnetic field.
Interestingly without magnetic field we have real exponents \cite{Karch:2009ph},
\begin{equation}
\delta_1=-2;~~~\delta_2=-3;\ 
\end{equation}
and the phase transition is continuous. Furthermore one can show that it is a third order phase transition \cite{Karch:2009ph}. Note also that the linearized equation (\ref{EOMZoom}) is valid for large enough $z$ (as the negative sign of the exponents suggests), while the zoomed in geometry (\ref{metricZ}) is valid for small $z$. 
Therefore our analysis is valid in an intermediate region of values for $z$ and $y$, which exists since one can always consider sufficiently small constants in equation (\ref{XI}). Note that the D7--brane embeddings are uniquely determined by specifying initial conditions $y(z_0)=0$ for ``Minkowski" embeddings and $z(y_0)=0$ for ``ball" embeddings. 
We will focus on the ``Minkowski" class of embeddings since the distance above the ball, specified by $z_0$, can be interpreted as a dynamical mass of the fundamental fields \cite{Mateos:2007vn} and acts as a natural order parameter for our phase transition. Now the scaling property of the equation of motion mentioned earlier suggests that if we rescale the initial condition by $z_0'=z_0/\mu$, the constants in equation (\ref{XI}) scale as:
\begin{equation}
{{\rm const}}_1'={\rm const}_1\mu^{\delta_1-1};~~~{\rm const}_2'={\rm const}_2\mu^{\delta_2-1};\ .\label{scalconst}
\end{equation}
In the field theory the critical embedding $y_*$ corresponds to a critical state characterized by some bare mass $m_*$ and fundamental condensate $c_*$. Our next step is to assume that the field theory parameters $m,\,c$ corresponding to the embeddings in the vicinity of the critical embedding depend analytically on the constants ${\rm const}_1$ and ${\rm const}_2$. 
This is a reasonable assumption since the region where the linearized equation (\ref{XI}) holds is away from the conical singularity of the critical embedding. 
Therefore we can expand
\begin{eqnarray} 
m-m_*&=&A_1\,{\rm const}_1+A_2\,{\rm const}_2+\dots\ ,\\
c-c_*&=&B_1\,{\rm const}_1+B_2\,{\rm const}_2+\dots\nonumber\ ,
\end{eqnarray}
where we have introduced new sets of constants $A_1,A_2,B_1,B_2$. Now using the scaling property (\ref{scalconst}) we can write:
\begin{eqnarray}
m'-m_*&=&A_1\,{\rm const}_1\, \mu^{\delta_1-1}+A_2\,{\rm const}_2\,\mu^{\delta_2-1}+\dots\ ,\\ \label{scalMC}
c'-c_*&=&B_1\,{\rm const}_1\,\mu^{\delta_1-1}+B_2\,{\rm const}_2\,\mu^{\delta_2-1}+\dots\nonumber\ .
\end{eqnarray}
Next, we eliminate the scaling parameter $\mu$ from equation (\ref{scalMC}) and obtain, to leading order,
\begin{eqnarray}
c-c_*=D_1(m-m_*)+D_2(m-m_*)^{\frac{\delta_2-1}{\delta_1-1}}+\dots\ ,
\end{eqnarray}
where we have omitted the prime superscript and defined another set of constants $D_1,D_2$. In our case $\delta_1=-2,\delta_2=-3$ and we find \cite{Karch:2009ph}:
\begin{equation}
c-c_*=D_1(m-m_*)+D_2(m-m_*)^{4/3}+\dots\ . \label{CvsM}
\end{equation}
Clearly, equation (\ref{CvsM}) suggests that the fundamental condensate of the theory is a continuous function of the bare mass with regular first derivative and divergent second derivative. 
Given that the condensate is a first derivative of the free energy, we conclude that the phase transition is of third order~\cite{Karch:2009ph}. 
\section{External magnetic field}
In this section we study the influence of an external magnetic field on the flavoured gauge theory. First we focus on the effect of the magnetic field on the confinement/deconfinement phase transition. For sufficiently strong magnetic fields we study the spontaneous breaking of the axial $U(1)$ $R$-symmetry. 
We supplement our studies with an analysis of the meson spectrum.  
\subsection{Introducing magnetic field}
In order to couple the fundamental fields to an external magnetic field we turn on a pure gauge $B$-field along two of the directions of the $S^3$ where the dual gauge theory lives. Unlike the flat case considered in ref. \cite{Filev:2007gb} a constant $B$-field is not the natural choice on $S^3$. It is instructive to write the metric of $S^3$ in terms of local tetrads:
\begin{equation}
d\Omega_3^{(1)\,2}={e^{(1)}}^2+{e^{(2)}}^2+{e^{(3)}}^2
\end{equation}
where the tetrads are defined by:
\begin{eqnarray}
e^{(1)}=R\,d\theta_1\ ,~~~e^{(2)}=R\,\sin\theta_1\,d\phi_1\ ,~~~e^{(3)}=R\,\cos\theta_1\,d\psi_1\ .
\end{eqnarray}
A natural choice for the $B$-field is:
\begin{equation}
B=H e^{(1)}\wedge e^{(2)}\ .
\end{equation}

It is easy to check that the pure gauge condition $B=dA$ is satisfied for $A=-HR^2\cos\alpha\,d\beta$. It is now straightforward to show that the DBI lagrangian from equation (\ref{lagr}) is modified to:
\begin{equation}
{\cal L} \propto u\cos^3\theta_3\left(1-\frac{R^4}{16u^4}\right)\sqrt{u^4\left(1-\frac{R^2}{4u^2}\right)^4+H^2R^4}\sqrt{1+u^2\theta_3'(u)^2}\ .\label{lagrB} 
\end{equation}
However, it turns out that with this ansatz for the $B$-field the equation of motion for the gauge field is not satisfied.\footnote{This fact was overlooked in a previous version of this paper as pointed out in ref.~\cite{Chunlen:2014zpa}.} Indeed, one can show that the variation of the DBI action (\ref{DBI}) with respect to the $A_{\phi_1}$ component of the gauge field is non-zero:
\begin{equation}
\frac{\delta S_{\rm DBI}}{\delta A_{\phi_1}} \neq 0\ .
\end{equation}
A possible resolution is to introduce an external seven form Ramond-Ramond flux $F_{(7)}$,~which couples to the Wess-Zumino action of the D7-brane through the term $\mu_7\int A_{(1)}\wedge F_{(7)}$. The required flux is then given by:
\begin{equation}\label{new-eq-A}
{F_{(7)}}_{t\,\theta_1\,\psi_1\, u\,\alpha\,\beta\,\gamma}=\frac{1}{\mu_7}\frac{\delta S_{\rm DBI}}{\delta A_{\phi_1}}\ ,
\end{equation}
where $\alpha,\beta$ and $\gamma$ parameterise the $S^3\subset S^5$ wrapped by the D7--brane and equation (\ref{new-eq-A}) is our new equation of motion for the gauge field, which is now satisfied. The physical meaning of the $F_{(7)}$ flux is the following: It represents the external electric currents needed to support the magnetic field on the three-sphere. Note that an honest top-down model would require taking into account the back reaction of this flux; in this paper, we take a semi-bottom-up approach and ignore its back reaction. 

Another feature of this ansatz is that the norm of the $B$-field diverges at the origin of AdS$_5$, which is problematic for the ``ball'' embeddings. In fact, this divergence is a consequence of the fact that the magnetised three-sphere shrinks at the origin and suggests that there is a magnetic monopole located there. A magnetic monopole in the eight dimensional world volume of the D7-brane is a four dimensional object, which in string theory is realised by the boundary of a stack of D5--branes ending on the D7--brane. In our semi-bottom-up approach we will not be interested in the exact details of how this D5--brane configuration is realised, and will ignore its back reaction, which is equivalent to ignoring the back reaction of the $F_{(7)}$ flux introduced above, since it is sourced naturally by the D5-branes.

Next, we address the question how the external magnetic field affects the phase transition described in section~1. To answer that question we need to solve numerically for the D7-brane embeddings and study the corresponding equation of state. However, at least for small magnetic field one would expect that the classification of the embeddings remains the same and there is still a critical embedding with a conical singularity at the origin of the $AdS_5$ space 
(represented by the shell of the ball at $u=R/2$) separating the two classes of embeddings. It is then natural to calculate the corresponding scaling exponents. After performing the change of coordinates (\ref{zooming}) and zooming into the vicinity of the critical embedding, we obtain the following Lagrangian
\begin{equation}
{\cal L}_{zoom}\propto R^2H z y^3\sqrt{1+y'^2}\ .
\end{equation}
Notice that the power of $z$ has changed and now we have the case $k=2,~~n=3$ in equation (\ref{lagrgen}). The critical exponents are then given by equation (\ref{deltapm}) and we obtain
\begin{equation}
\delta_{\pm}=-\frac{3}{2}\pm i \frac{\sqrt{7}}{2}
\end{equation}
Remarkably the critical exponents are complex\footnote{In fact these are the same scaling exponents as in the finite temperature case when the singular shell separating the two classes of embeddings is an event horizon.}. Therefore the system exhibits discrete self-similar behaviour and the equation of state in the $c$ versus $m$ plane has a multi-valued behavior near the critical state $(\tilde m_*,\tilde c_*)$, seeded by a spiral structure. 
Thus we expect that when we move away from the $H=0$ point, the third order phase transition becomes a first order phase transition. 

Following the same logic as for the continuous case considered in section 2.2, we consider approaching the critical embedding by scaling the initial conditions, namely $z_0'=z_0/\mu$ for ``ball" embeddings and $y_0'=y_0/\mu$ for ``Minkowski" embeddings. Then the corresponding values of the parameters $\tilde m$ and $\tilde c$ scale as \cite{Frolov:2006tc, Filev:2008xt}:
\begin{equation}
\begin{pmatrix} \tilde m'-\tilde m^*\\  \tilde c'-\tilde c^* \end{pmatrix}=\frac{1}{\mu^{5/2}}M\begin{pmatrix}\cos{\left(\frac{\sqrt{7}}{2}\ln\mu\right)} & \sin{\left(\frac{\sqrt{7}}{2}\ln\mu\right)}\\ -\sin{\left(\frac{\sqrt{7}}{2}\ln\mu\right)}& \cos{\left(\frac{\sqrt{7}}{2}\ln\mu\right)}\end{pmatrix}M^{-1}\begin{pmatrix} \tilde m-\tilde m^*\\  \tilde c-\tilde c^* \end{pmatrix}\ , \label{mcscalling}
\end{equation}
where $M$ is a constant non-singular $2\times2$ matrix. Equation (\ref{mcscalling}) can be checked numerically by solving (numerically) the equations of motions derived form equation (\ref{lagrB}). In fact, it turns out to be more convenient to change variables, i.e.,
\begin{equation}
\rho=u\,\cos\theta_3\ ,~~~L=u\,\sin\theta_3\ .
\end{equation} 
The Lagrangian (\ref{lagrB}) becomes
\begin{equation}
{\cal L}\propto \tilde\rho^3 \left(1-\frac{1}{16(\tilde\rho^2+\tilde L^2)^2}\right) \sqrt{\left(1 - \frac{1}{4(\tilde\rho^2+\tilde L^2)}\right)^4+\frac{H^2}{(\tilde\rho^2+\tilde L^2)^2}} 
\sqrt{1+\tilde L'^2}, \label{LagrBLrho}
\end{equation} 
where we used the dimensionless coordinates $\tilde L=L/R$ and $\tilde\rho=\rho/R$. The $B$-field introduces a new logarithmic divergence that can be regulated by adding an additional counterterm to the boundary, ${\cal L_{H}}\propto B^2/4\log(\rho_{max}/R)$. Note that his counterterm is independent of the bare mass and therefore does not modify the fundamental condensate of the theory. 
The parameters $\tilde m$ and $\tilde c$ are obtained by expanding $\tilde L(\tilde \rho)$ for large $\tilde\rho$,
\begin{equation}
\tilde L=\tilde m+\frac{\tilde c_1}{\tilde\rho^2}-\frac{\tilde m}{2\tilde\rho^2}\log\tilde\rho+\dots\ .
\end{equation}
The condensate is calculated using equation (\ref{condcal}), where the dimensionful parameters $m,\,c$ and $c_1$ are used. 

\begin{figure}[htbp] 
   \centering
   \includegraphics[width=3.2in]{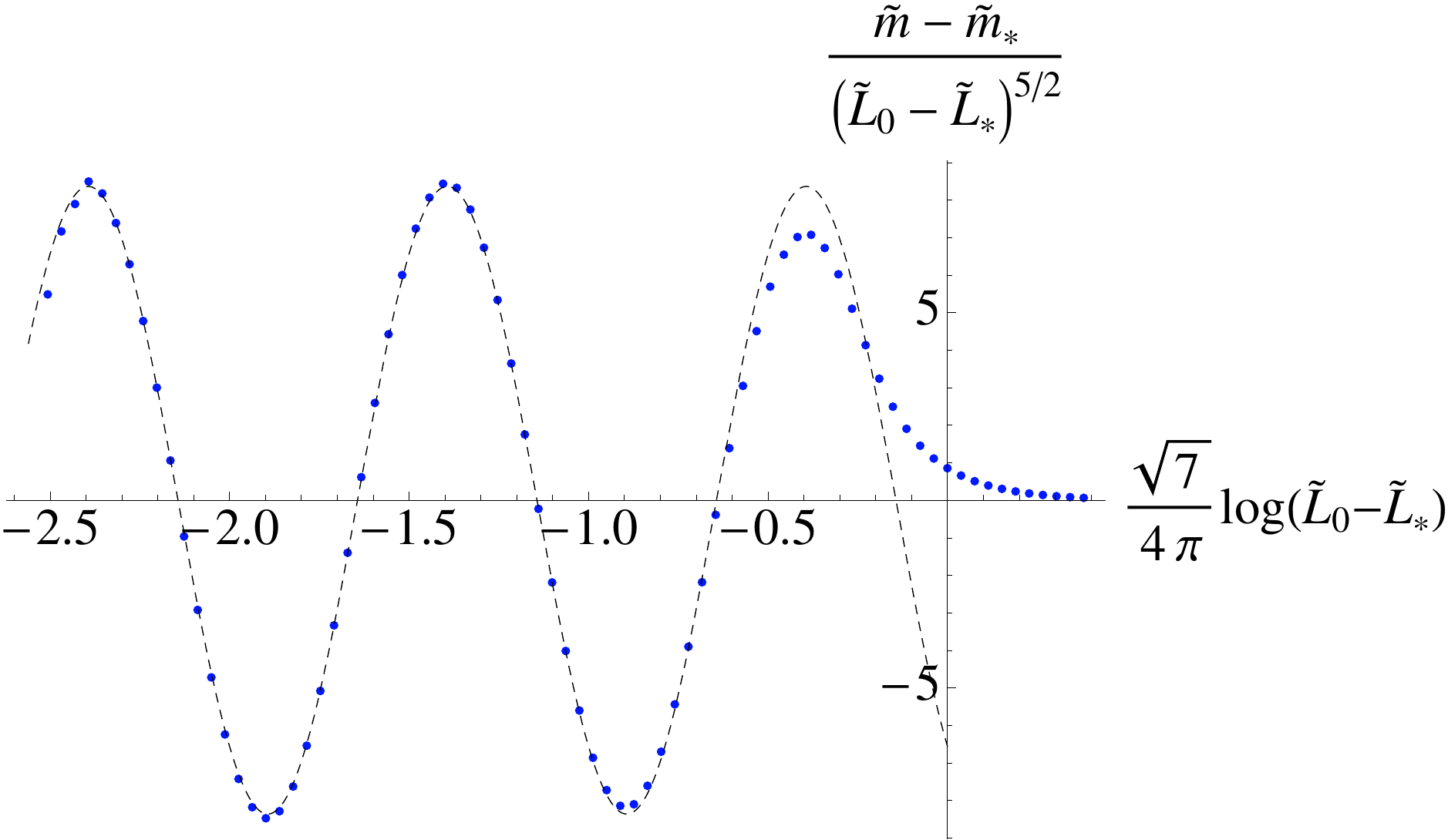} 
      \includegraphics[width=3.2in]{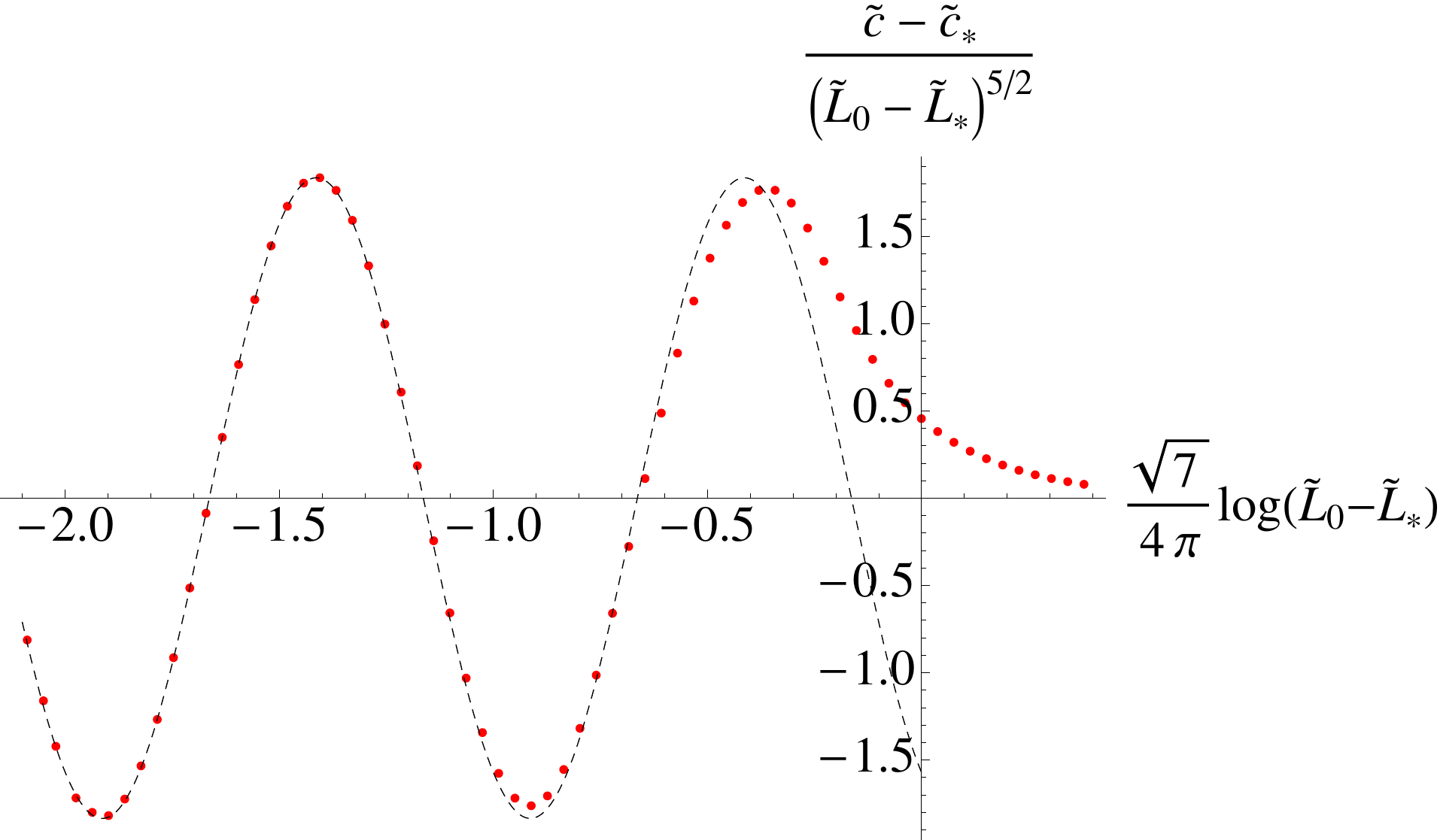}
      \caption{\small Plots representing the discrete self-similar structure of the theory near the critical state $(\tilde m_*,\tilde c_*)$. The states closer to the critical one are to the left of the horizontal axis. The dashed curves represent fits with trigonometric functions of unit period.}
   \label{fig:fig4}
\end{figure}

We now proceed to study numerically the self-similar structure of the theory near the critical state $(\tilde m_*,\tilde c_*)$ by approaching from the ``Minkowski'' class of embeddings. ``Minkowski'' embeddings are uniquely determined by specifying the initial value of $\tilde L$, namely, $\tilde L_0=\tilde L(0)$. The critical embedding corresponds to $\tilde L_0=\tilde L_*=1/2$. 
We can consider some initial value $L_0$ close to $L_*$ and then scale $L_0'=L_0/\mu$. In this way we can trade the scaling parameter $\mu$ in equation (\ref{mcscalling}) for $L_0/L_0'$ and study $m$ and $c$ as functions of $L_0$ for fixed $\tilde m',\tilde c'$ and $\tilde L_0'$. 
This suggests that if we plot the quantities $({\tilde m-\tilde m_*})/{(\tilde L_0-\tilde L_*)^{5/2}}$ and $({\tilde c-\tilde c_*})/{(\tilde L_0-\tilde L_*)^{5/2}}$ as functions of $(\sqrt{7}/4\pi)\log(\tilde L_0-\tilde L_*)$, we should obtain trigonometric functions of unit period. 
In figure \ref{fig:fig4} we have presented our numerical results for $H=0.3$. The dashed curves represent fits with trigonometric functions of unit period. One can see the excellent agreement as one explores states closer to the critical one (shifting toward negative values of $(\sqrt{7}/4\pi)\log(\tilde L_0-\tilde L_*)$).

\subsection{Equation of state and phase diagram}
In this subsection we analyze the dependence of the fundamental condensate on the bare mass. To this end we numerically solve the equation of motion for $\tilde L(\tilde\rho)$ obtained from equation (\ref{LagrBLrho}) and extract the parameters $\tilde m$ and $\tilde c$ from the asymptotics of the solution at large $\tilde\rho$. 
\begin{figure}[h] 
   \centering
   \includegraphics[width=3.0in]{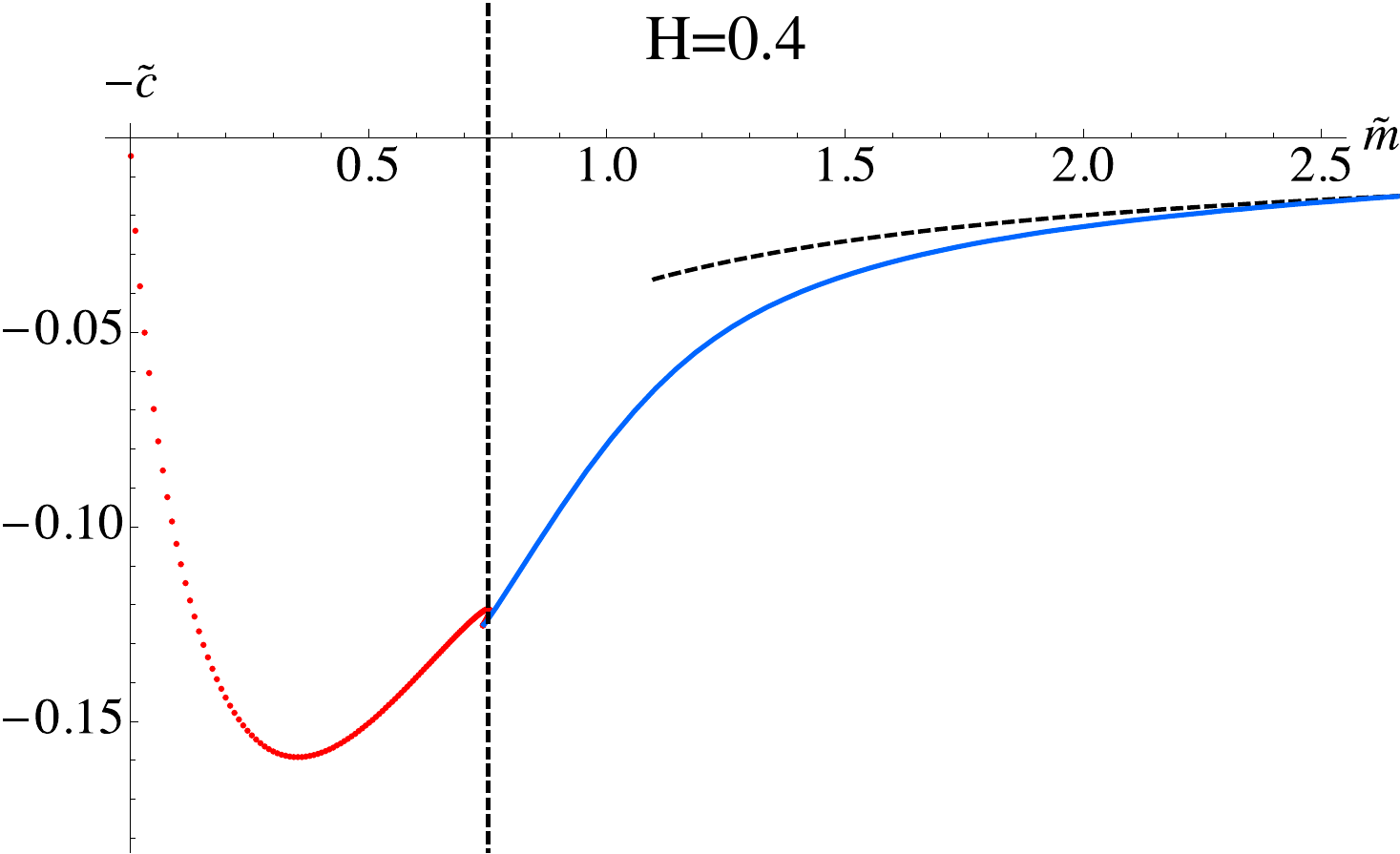} 
      \includegraphics[width=3.0in]{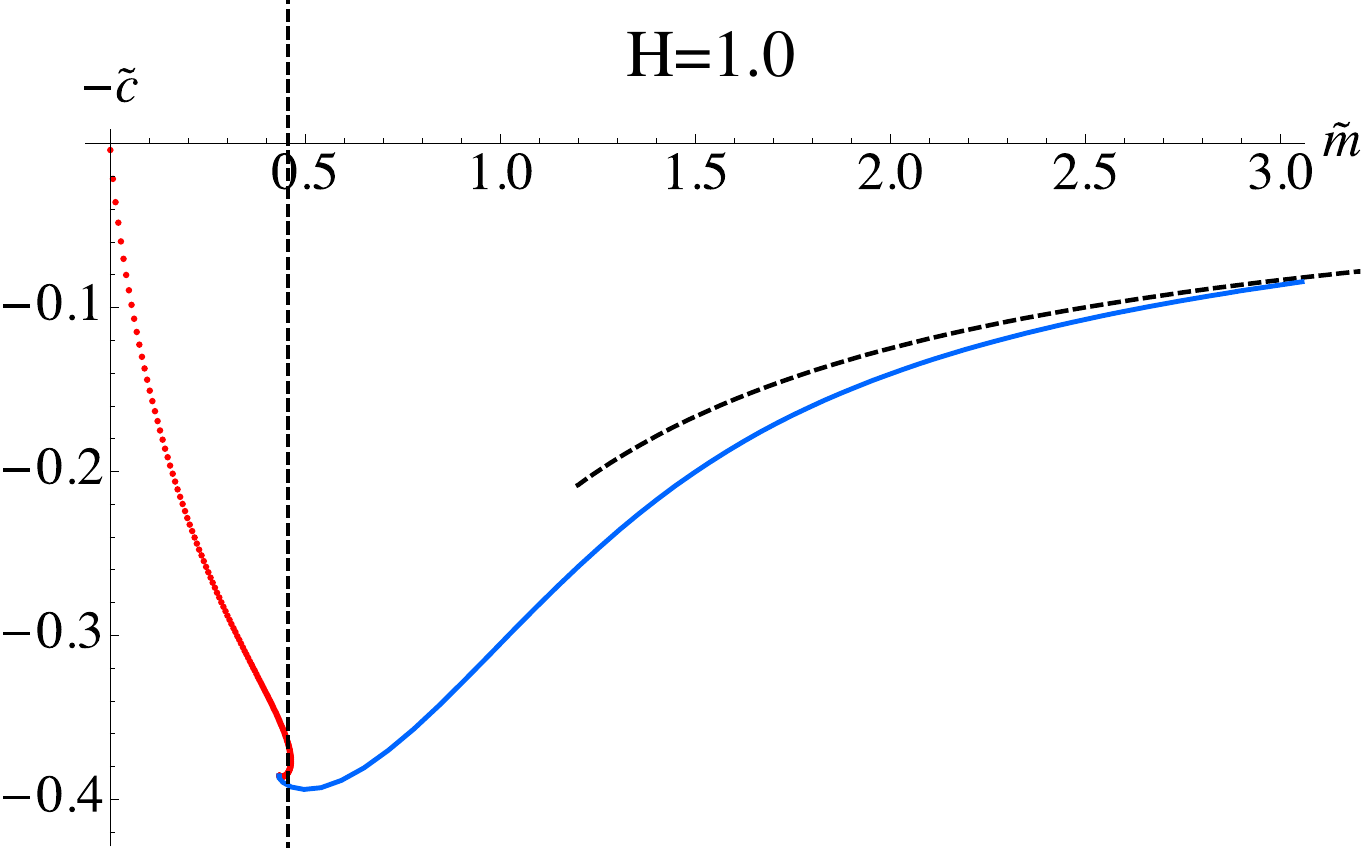} \\
      \includegraphics[width=3.0in]{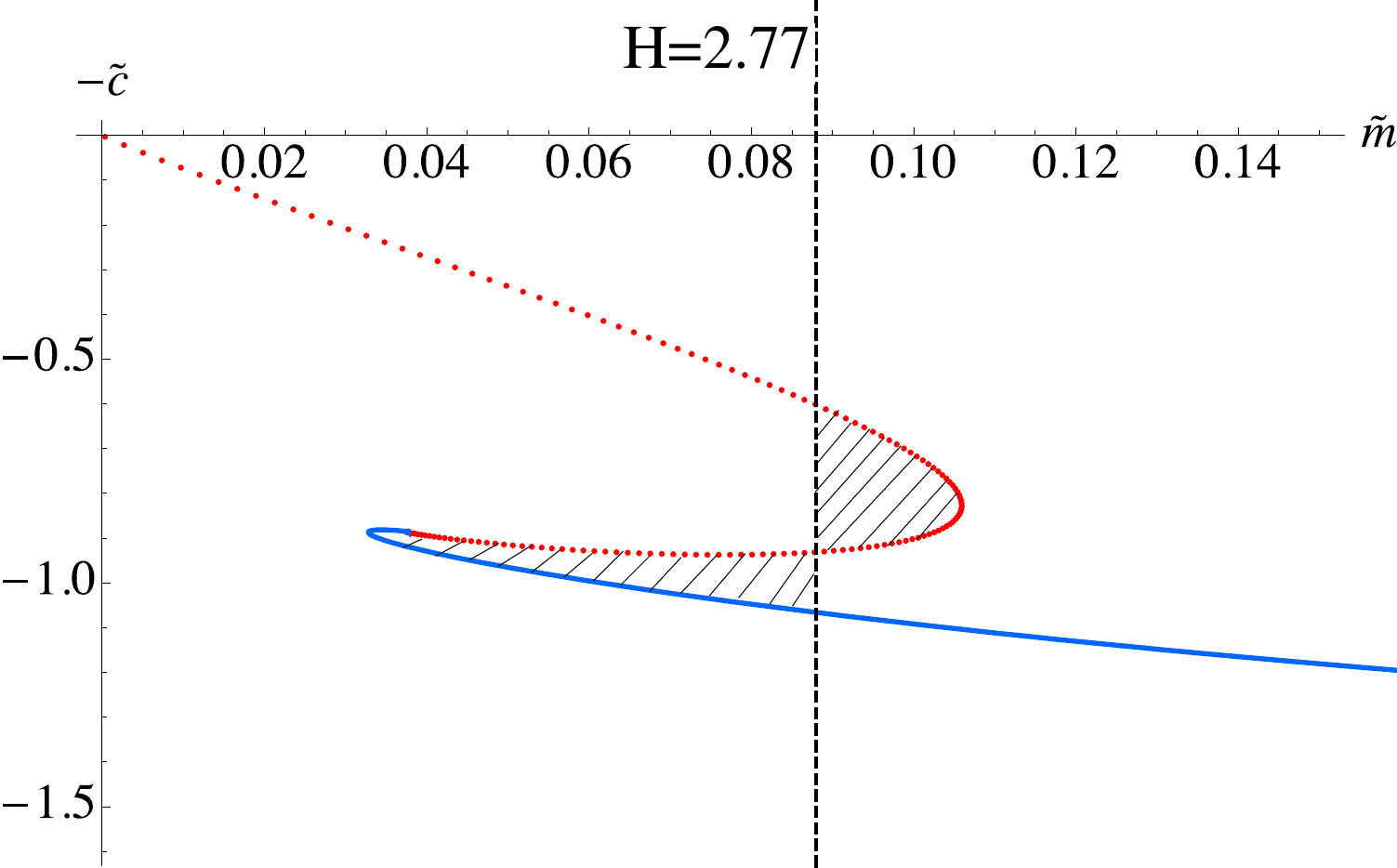}
            \includegraphics[width=3.0in]{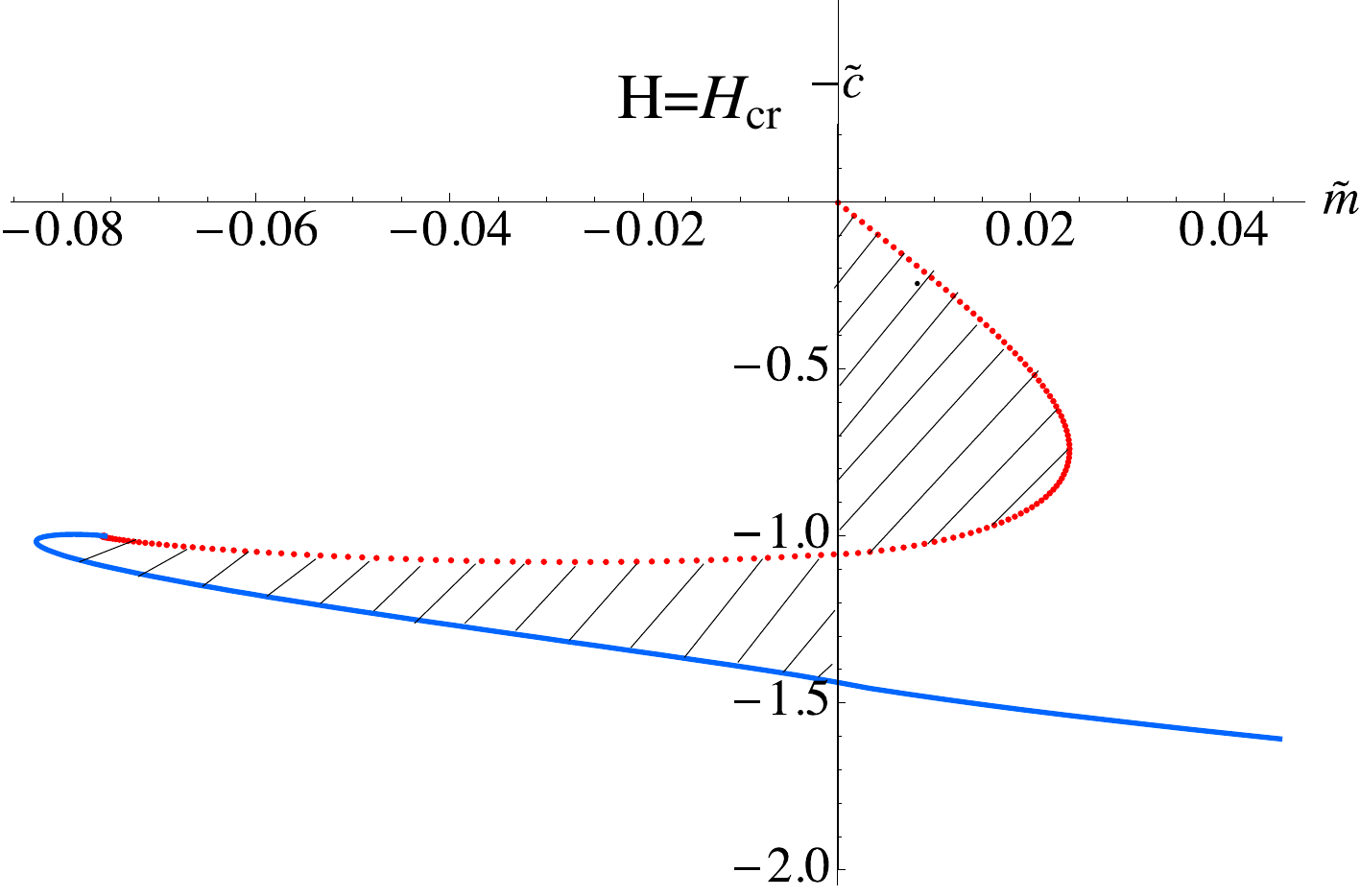}
 \caption{\small Plots of the condensate as a function of the bare mass for different values of the magnetic field. The black dashed curves represent a $1/\tilde m$ fit. At $H=H_{\rm cr}\approx 3.78$ the critical mass vanishes. Beyond this point the theory is in a confined phase with broken chiral symmetry and the zero bare mass state has a non-zero negative condensate.}
   \label{fig:fig5}
\end{figure}

In figure \ref{fig:fig5} we present our results for four different values of the magnetic field. As one can see, the effect of the magnetic field is to decrease the critical mass at which the confinement/deconfinement phase transition takes place. The dashed fitting curve in the figure represents a $1/\tilde m$ fit for large $\tilde m$. 
Moreover, for sufficiently large magnetic field $H=H_{\rm cr}\approx 3.78$ the critical mass vanishes and the transition happens at zero bare mass. Beyond this point the phase transition seizes to exist and at vanishing bare mass the stable vacuum has non-vanishing negative condensate spontaneously breaking the axial $U(1)$ R-symmetry. In our setup this breaking is analogous to the chiral symmetry breaking in the QCD vacuum. 
Thus we interpret this result as a manifestation of magnetic catalysis of chiral symmetry breaking in the flavoured gauge theory on $S^3$. 

The shaded regions in figure \ref{fig:fig5} demonstrate the equal-area law which can be used to determine the critical mass. The condensate is a derivative of the free energy with respect to the bare mass and in this setup one can show that, by integrating numerically the condensate as a function of the bare mass, one can obtain the free energy  (up to a additive constant) calculated by regularizing the euclidean on-shell action.
This is why the use of the equal area law is justified. Note that from the plots in figure \ref{fig:fig5} one can observe that the disrete self-similar regime of the theory, analyzed in the previous section, is thermodynamically unstable. 
Our studies of the meson spectrum confirm that it is also unstable under quantum fluctuations. 

\begin{figure}[htbp] 
   \centering
   \includegraphics[width=4.2in]{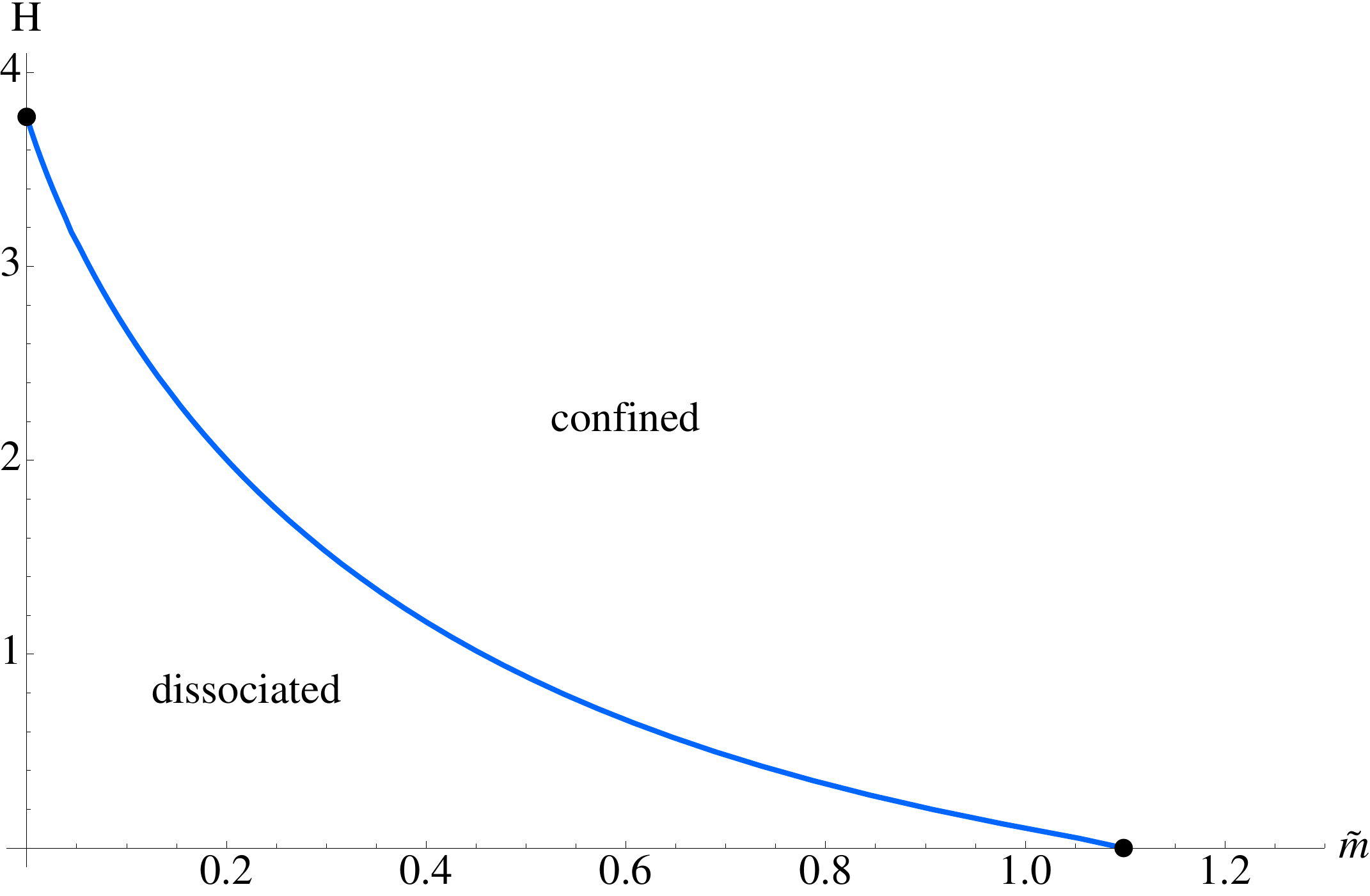} 
   \caption{\small A plot of the phase diagram of the theory. For finite magnetic fields the phase transition is of first order, while at vanishing magnetic field the phase transition is of third order. For sufficiently large magnetic fields the phase transition seizes to exist and the theory is in a confined phase.}
   \label{fig:fig6}
\end{figure}

We proceed by obtaining the phase diagram of the theory. To this end we numerically generate the fundamental condensate as a function of the bare mass parameter for different values of the magnetic field. 
Then we integrate numerically to obtain the free energy and to determine the critical mass of the transition. 
Our findings are presented in figure \ref{fig:fig6}. The phase diagram is similar to the one studied in ref. \cite{Albash:2007bk,Erdmenger:2007bn}. The analogue of the temperature in our scenario is the Casimir energy of the theory on $S^3$, which dissociates the mesons states. 
In our case however, the phase transition takes place at zero temperature and is thus a quantum phase transition. 
An interesting property of the phase diagram is that the first order phase transition for finite magnetic fields becomes a third order phase transition for vanishing magnetic field. 
One can also see that for sufficiently large magnetic field there is no phase transition at all and the theory is in a confined phase.

\subsection{Meson spectra with external magnetic field}
Another focus of the present work is to study the effect of an external magnetic field on the spectra of mesons in global AdS. The presence of a magnetic field 
manifests itself, as expected, by Zeeman splitting and level crossing. This will be studied in detail below. 
We explore the dependence of the spectra on the bare mass parameter for a large range of values of the magnetic field.
\subsubsection{Derivation of the fluctuation equations}
To investigate the spectrum of light mesons, we study quadratic fluctuations of the D7--brane embedding along the transverse coordinates $L,\phi_3$, 
which we expand in the following way:
\begin{equation}
L(\rho)=L_0(\rho) + 2 \pi \alpha' \chi(\rho), \qquad \phi_3(\rho) = 0 + 2 \pi \alpha' \Phi(\rho),
\end{equation}
where $L_0(\rho)$ is the classical D7--brane embedding.
The discussion closely follows earlier work reported in \cite{Filev:2007gb,Albash:2007bk,Erdmenger:2007bn,Filev:2009xp}.
Recall the induced metric of the D7--branes in ($L$, $\rho$) coordinates,
\begin{align}
ds^2_{\mathrm{D7}}&= -\frac{\rho^2 +L^2}{R^2}\left(1+\frac{R^2}{4(\rho^2+L^2)}\right)^2 d \tau^2 + \frac{\rho^2 +L^2}{R^2}\left(1-\frac{R^2}{4(\rho^2+L^2)}\right)^2
d \Omega_3^{(1)2} \nonumber  \cr
& \quad + \frac{R^2}{\rho^2 +L^2}\left[ \left( 1+L'(\rho)^2 \right) d\rho^2 + \rho^2 d \Omega_3^{(2)2}\right].
\end{align}
The equations of motion can be obtained straightforwardly from the appropriate DBI and WZ actions:
\begin{align}
S&=S_{\mathrm{DBI}}+S_{\mathrm{WZ}},\cr
S_{\mathrm{DBI}}&=-N_f\mu_{7}\int\limits_{{\cal M}_{8}}d^{8}\xi e^{-\Phi}[-{\rm det}(E_{ab}+2\pi\alpha' F_{ab})]^{1/2},\cr
S_{\mathrm{WZ}}&=\frac{(2\pi\alpha')^2}{2}\mu_{7}\int{F_{(2)}\wedge F_{(2)}\wedge P[C_{(4)}]}+(2\pi\alpha')\mu_{7}\int F_{(2)}\wedge B_{(2)}\wedge P[\tilde{C}_{(4)}]+\nonumber\\
&+(2\pi\alpha')\mu_{7}\int A_{(1)}\wedge F_{(7)}
\label{WZ},\cr
\end{align}
where $P[C_{(4)}]$ is the pull-back of the 4-form potential sourced by the stack of $N_c$ D3--branes, $P[\tilde{C}_{(4)}]$ is the pull-back of its magnetic dual and $F_{(7)}$ is the external flux that we introduced.
Expanding $E_{ab}$ to second order in $\alpha'$, we have
\begin{equation}
 E_{ab}= E^0_{ab} + 2 \pi \alpha' E^1_{ab} + (2 \pi \alpha')^2 E^2_{ab},
\end{equation}
where 
\begin{align*}
E^0_{ab}&= g_{ab}(L_0(\rho)) + B_{ab},\cr
E^1_{ab}&= G_{LL}  L_0'(\rho) \left(\partial_a \chi \delta_b^{\rho} + \partial_b \chi \delta_a^{\rho}\right) + \left(\partial_{L_0} g_{ab}\right) \chi
,\cr
E^2_{ab}&= G_{LL} \left( \partial_a \chi \partial_b \chi + L_0^2  \partial_a \Phi \partial_b \Phi \right) + \left(\partial_{L_0} G_{LL}  L_0' \right) 
\left(\partial_a \chi \delta_b^{\rho} + \partial_b \chi \delta_a^{\rho}\right) \chi + \frac{1}{2} \partial^2_{L_0} G_{ab} \chi^2,
\end{align*}
where $g_{ab}$ and $B_{ab}$ are the induced metric and B-field on the D7--brane world volume and the prime stands for derivation w.r.t. $\rho$.  
Splitting $\left(E^0_{ab}\right)^{-1}$ into symmetric and anti-symmetric parts, i.e., $\left(E^0_{ab}\right)^{-1}=: S^{ab} + J^{ab}$, where 
\begin{align*}
S^{ab}&= \mathrm{diag} \left\{ -g^{-1}_{tt}, \frac{g_{33}}{g^2_{33}+H^2}, \frac{g_{33}}{g^2_{33}+H^2}, g^{-1}_{33}, g^{-1}_{\rho \rho}, g^{-1}_{\theta_2 \theta_2}, 
g^{-1}_{\phi_2 \phi_2}, g^{-1}_{\psi_2 \psi_2} \right\},\cr
J^{ab}&=\frac{H}{g^2_{33}+H^2}\left( \delta_2^a \delta_1^b - \delta_2^b \delta_1^a \right),\cr
g_{tt}&=-\frac{\rho^2 +L_0^2}{R^2}\left(1+\frac{R^2}{4(\rho^2+L_0^2)}\right)^2,\; g_{33}=  \frac{\rho^2 +L_0^2}{R^2}\left(1-\frac{R^2}{4(\rho^2+L_0^2)}\right)^2, \cr
g_{\rho \rho}&= \frac{R^2}{\rho^2 +L_0^2}\left(1+L_0'(\rho)^2\right). 
\end{align*}
It was demonstrated in refs.~\cite{Filev:2007gb, Arean:2005ar} that the effect of the magnetic field regarding the equations of motion is to couple the scalar and vector modes; namely,
$\Phi$ will couple to the $A_0$ and $A_3$ components of the gauge field, while the $\chi$ fluctuations will mix with the $A_1$ and $A_2$ in the presence of a 
magnetic field. This can be understood from the fact that our ansatz for the external magnetic field breaks part of the remaining 3+1 dimensional symmetry. Note that Lorentz invariance is broken 
already, cf.~(\ref{Globu}).\\
With this in mind, we arrive at the following expressions for the relevant terms in the Lagrangian\footnote{Note that the $\mathcal{O}(\alpha')$ part of the action is equal to a total derivative, as it should be, since we are expanding near a local extremum of the action, because in the presence of the external flux $F_{(7)}$ the equations of motion for the classical embedding are satisfied.}:
\begin{align}
\frac{\mathcal{L}_{\chi\chi}^{(2)}}{\sqrt{g_{S^3}}}&=\frac{1}{2} g(\rho) G_{LL}\frac{S^{t t}}{1+L_0'^2}\partial_{t}\chi\partial_{t}\chi +\frac{1}{2} g(\rho) G_{LL}\frac{S^{\rho \rho }}{1+L_0'^2}\partial_{\rho}\chi\partial_{\rho}\chi \\\nonumber
& \quad +\frac{1}{2}g(\rho) \left[\partial_{L_0}\left(\partial_{L_0} \log g(\rho)\right) -\frac{L_0'}{1+L_0'^2} \partial_{\rho}\left( \partial_{L_0} \log g(\rho)\right)\right]\chi^2\ \label{LagChi},\cr
\frac{\mathcal{L}_{\Phi\Phi}^{(2)}}{\sqrt{g_{S^3}}}&= \frac{1}{2} g(\rho) G_{\phi_3 \phi_3} \left( S^{t t}\partial_{t}\Phi\partial_{t}\Phi + S^{\rho \rho}\partial_{\rho}\Phi\partial_{\rho}\Phi\right),\cr
\frac{\mathcal{L}_{AA}^{(2)}}{\sqrt{g_{S^3}}}&=-\frac{1}{4} g(\rho) S^{aa'} S^{bb'} F_{ab}F_{a'b'},\cr
\frac{\mathcal{L}_{\Phi A}^{(2)}}{\sqrt{g_{S^3}}}&= - H \left(\partial_{\rho} K(\rho)\right) \Phi F_{03}.
\end{align}
The metric component $G_{\phi_3 \phi_3}$ and the functions $g(\rho)$ (the Lagrangian density) and $K(\rho)$ are given by
\begin{align}
G_{\phi_3 \phi_3}&=\frac{R^2 L_0^2}{\rho^2 +L_0^2},\cr
g(\rho)&:=\frac{ \sqrt{-{\rm det} E^0_{ab}}}{\sqrt{g_{S^3}}}=\rho^3 \left(1-\frac{R^4}{16(\rho^2+L_0^2)^2}\right) \sqrt{\left(1 - \frac{R^2}{4(\rho^2+L_0^2)}\right)^4+\frac{H^2R^4}{(\rho^2+L_0^2)^2}} 
\sqrt{1+L_0'(\rho)^2}, \cr
K(\rho)&= \frac{R^4 \rho^4}{(\rho^2+L_0^2)^2}.\\
\sqrt{g_{S^3}}&=\sin\alpha\cos\alpha\ ,
\end{align}
Now it is straightforward to obtain the equations of motion for the fluctuations from the usual Euler-Lagrange procedure, yielding
\begin{eqnarray}
&&\frac{1}{g(\rho)}\partial_{\rho}\left(\frac{g(\rho)\partial_{\rho}\chi}{(1+L_0^{\prime 2})^2}\right)+\frac{R^4 \omega^2 \chi}{(1+L_0^{\prime 2})(\rho^2+L_0^2)^2\left(1+\frac{R^2}{4(\rho^2+L_0^2)}\right)^2} \nonumber\\
&& \qquad \qquad \qquad -\left[\partial_{L_0}\left(\partial_{L_0} \log g(\rho)\right) -\frac{L_0'}{1+{L_0'}^2} \partial_{\rho}\left( \partial_{L_0} \log g(\rho)\right)\right]\chi=0,\label{eomL}\\
&&\frac{1}{g(\rho)}\partial_{\rho}\left(\frac{g(\rho)L_0^2\partial_{\rho}\Phi}{1+L_0^{\prime 2}}\right)+\frac{L_0^2 R^4 \omega^2 \Phi}{(\rho^2+L_0^2)^2\left(1+\frac{R^2}{4(\rho^2+L_0^2)}\right)^2}-\frac{H\partial_{\rho}K}{g(\rho)}F_{03}=0 ,\label{eomPhi}\\
&&\frac{1}{g(\rho)}\partial_{\rho}\left(\frac{g(\rho)\partial_{\rho}F_{03}}{(1+L_0^{\prime 2})\left(1-\frac{R^2}{4(\rho^2+L_0^2)}\right)^2}\right)+\frac{R^4}{(\rho^2+L_0^2)^2\left(1-\frac{R^4}{16(\rho^2+L_0^2)^2}\right)^2} \omega^2 F_{03}-\frac{H\partial_{\rho}K}{g(\rho)}\omega^2 \Phi=0,\nonumber\\ \label{eomF}
\end{eqnarray}
where $F_{03}=\partial_0 A_3-\partial_3 A_0$. We have assumed a plane wave ansatz for the fluctuations of the form $\delta X(t,\rho) = e^{-i \omega t} \delta X(\rho)$.
A few remarks are in order: We will be interested in investigating the spectrum of ``pions" which correspond to fluctuations along $\phi_3$. Therefore we will only consider the gauge field components 
$A_0$ and $A_3$ which couple to $\Phi$ and solve the corresponding coupled system of differential equations. Similar equations of motion were obtained and discussed e.g. in \cite{Filev:2007gb} for the case of $AdS_5 \times S^5$.
In order to be able to derive an equation in terms of $F_{03}$, we have to define $F_{03}\equiv \partial_{\tau} A_3$, setting $\partial_3 A_0=0$. This is due to the broken 
$SO(1,1)$ symmetry of the Minkowski part of (\ref{Globu}). Moreover, since we assume $\partial_i \chi=0$, the $\chi$ equations decouple from those for the gauge field components 
$A_1$ and $A_2$. \\
In the following sections, we will present numerical solutions to the equations of motion (\ref{eomL})-(\ref{eomF}) which were obtained employing a shooting technique  
in Mathematica. The general strategy is to start with appropriate initial conditions and shoot towards the boundary where we are interested in finding those solutions
that display the expected fall-off behaviour in the UV.

\subsubsection{Fluctuations along $L$}
\begin{figure}[htbp] 
   \centering
   \includegraphics[width=4.2in]{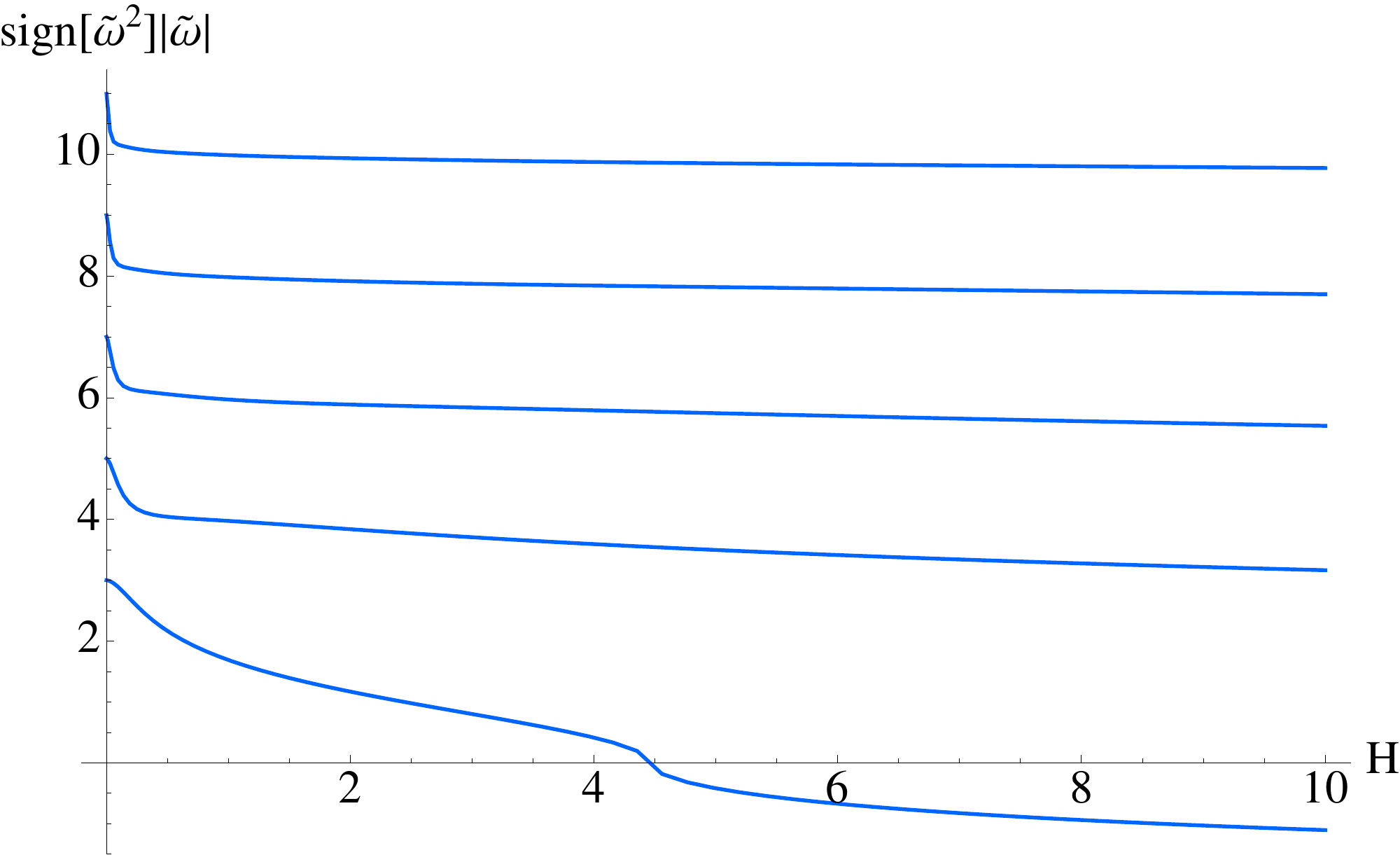} 
 \caption{\small The spectrum of $\chi$ fluctuations for zero bare mass (trivial embedding) vs. the magnetic field $H$.}
   \label{fig:zeromass}
\end{figure}

Here, we will present our numerical study of the meson spectrum associated with fluctuations along $L$. 
As a first step, we obtain the spectrum at zero bare mass, i.e., for the trivial embedding wrapping the $S^3$ within the $S^5$.\\
\noindent
{\bf Zero bare mass.} Figure \ref{fig:zeromass} shows the dependence of $\omega_i$ on the strength of the magnetic field $H$ for the lightest four meson states, $i=1,\ldots,4$.. 
We plot ${\rm sgn}(\omega_i^2) |\omega_i|$ which becomes negative when $\omega_i$ becomes imaginary (i.e., the state becomes tachyonic).
 For $H \rightarrow 0$, the spectrum is discrete and equidistant with eigenfrequencies given by \cite{Erdmenger:2010zm}
\begin{equation}
\omega = (2n+3) \frac{1}{R}. 
\end{equation}
Increasing the magnetic field, we observe that the lightest meson state becomes tachyonic, i.e., $\omega_1$ becomes imaginary, for some $H_{\rm cr}>H_{\ast}$.
This indicates the existence of a metastable phase between $H_{\ast}$ and $H_{\rm cr}$, which becomes unstable for even larger $H$.
Note that only the scalar modes condense, i.e become unstable, whereas the vector modes remain stable. This is not in contradiction with the proposed condensation
of the charged $\rho$ meson discussed in \cite{Chernodub:2010qx}, since we consider only one type of flavour and therefore all meson modes are neutral. \\
\noindent
{\bf Nonzero quark mass.} We proceed by investigating the spectrum of fluctuations along L as a function of the quark mass parameter $\tilde{m}$. We again solve numerically the equation of motion for $\chi$, eq. (\ref{eomL}), for the two classes of embeddings, ``Minkowski" and ``ball" embeddings.

\begin{figure}[h] 
   \centering
   \includegraphics[width=3.2in]{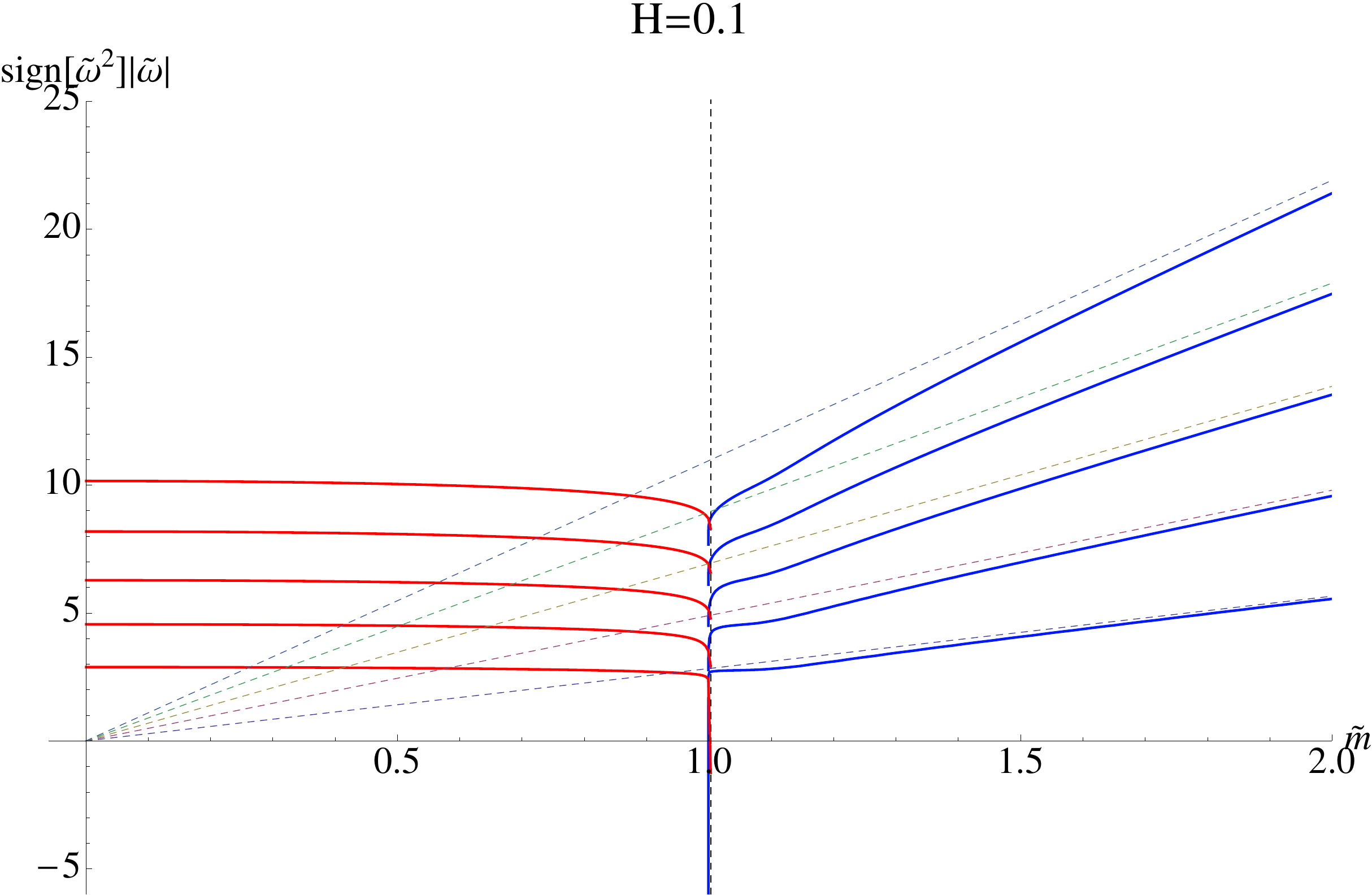} 
   \includegraphics[width=3.2in]{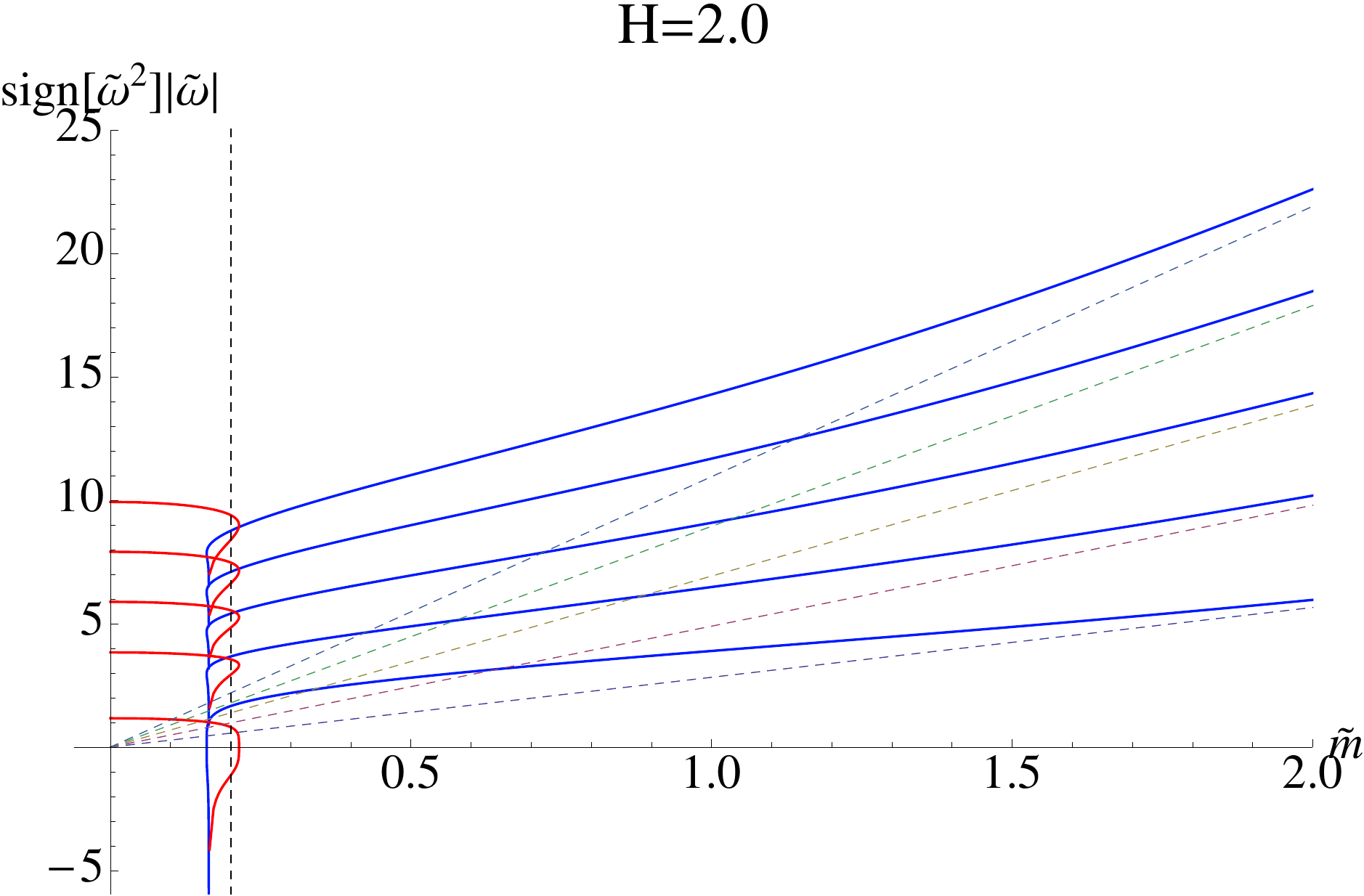} 
 \caption{\small The spectrum of $\chi$ fluctuations vs. bare quark mass $\tilde{m}$ for intermediate values of $H$, $H=0.1$ for the left plot and $H=2.0$ for the right one. The vertical dashed line indicates the location of the critical bare mass
       $\tilde{m}_{\ast}$, while the coloured dashed lines correspond to the spectrum of pure $AdS_5 \times S^5$. The red curves correspond to the ball embeddings and 
        the blue curves represent the Minkowski embeddings; the phase transition happens for the critical embedding separating the two classes.}
   \label{fig:SpectrumChi}
\end{figure}

For intermediate values of the magnetic field we present our results in figure \ref{fig:SpectrumChi}. The dashed lines in the figure represent the spectrum of the theory on $\IR^{1,3}$ without external magnetic field, studied in ref.~\cite{Kruczenski:2003be},
\begin{equation}
\omega = \frac{2m}{R^2}\sqrt{(n+1)(n+2)}, \label{spectflatcase}
\end{equation}
As expected, at bare masses larger than the energy scales set but the magnetic field and the Casimir energy, the spectrum is well described by equation (\ref{spectflatcase}).
As one approaches the phase transition, the spectrum becomes multivalued, with ``competing'' confined and deconfined phases. At the phase transition the spectrum has a finite jump. It is also interesting to see that close to the critical state, where the theory has a discrete self-similar structure, the spectrum becomes tachyonic, 
which suggests that the self-similar regime is unstable under quantum fluctuations and cannot be realized by ``supercooling" (i.e., it is not meta-stable). In the deconfined space the spectrum remains discrete because the theory is in a box.

\begin{figure}[h] 
   \centering
   \includegraphics[width=4.2in]{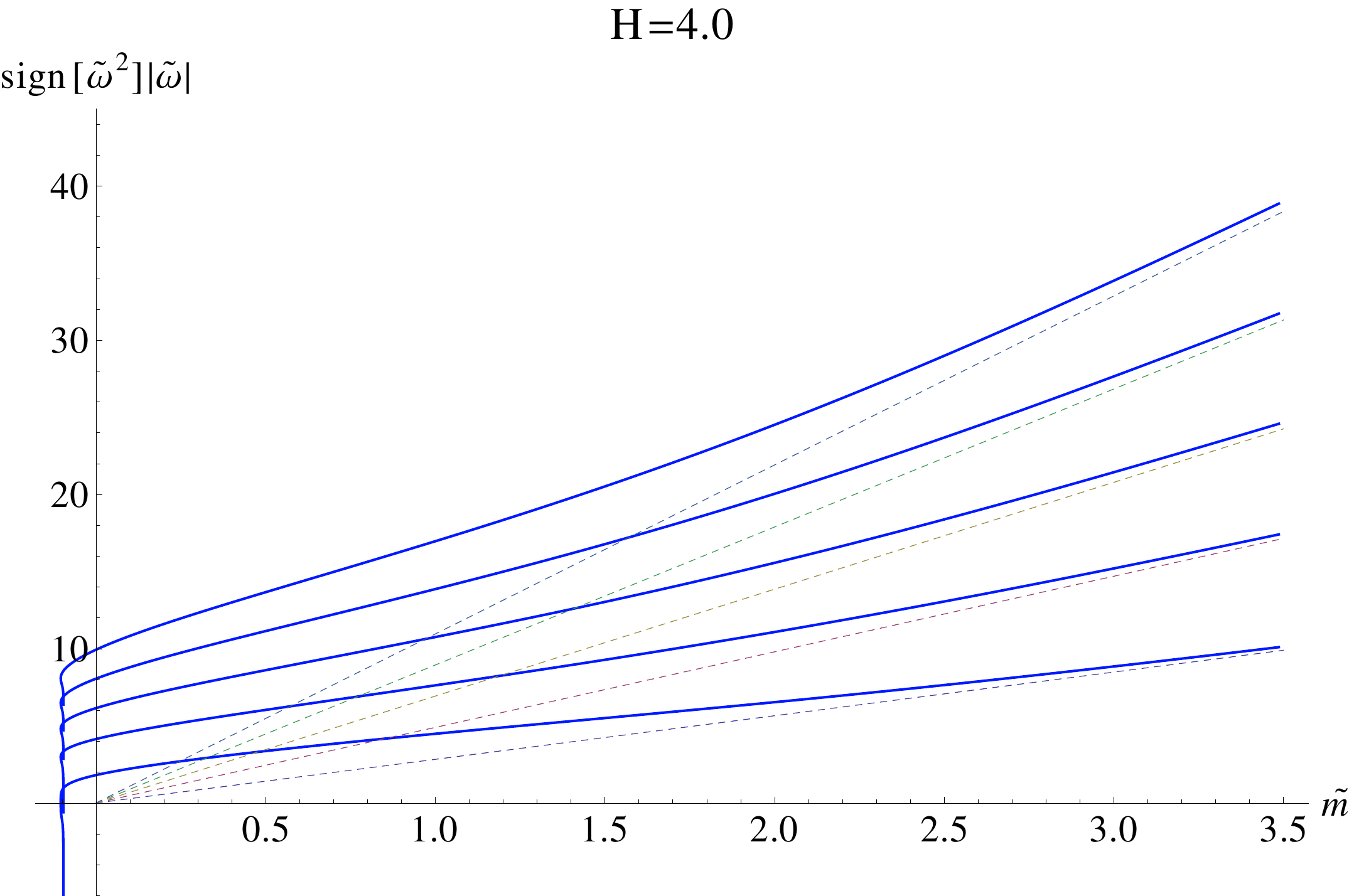} 
 \caption{\small A plot of the spectrum of fluctuations along $L$ for $H=4.0>H_*$. One can see that the positive $\tilde m$ branch is stable. }
   \label{fig:SpectrumChiH4}
\end{figure}

For magnetic fields above criticality, $(H>H_*\approx 3.78)$, the only stable phase is the confined phase. The corresponding spectrum is presented in figure \ref{fig:SpectrumChiH4}. As one can see the positive $\tilde m$ branch is stable.

\subsubsection{Fluctuations along $\phi_3$}

The spectrum of scalar mesons associated with fluctuations along $\phi_3$ is particularly interesting because it possesses a mode analogous to the $\eta'$-meson of large $N_c$ QCD. 
The analogy comes form the fact that rotational symmetry along $\phi_3$, present when $L=0$, corresponds to the anomalous axial $U(1)$ R-symmetry, which is restored at large $N_c$. Spontaneous breaking of this symmetry leads to a massless Goldstone boson, 
the analogue of $\eta'$ in QCD. In the gravity setup the repulsive potential due to the $B$-field decreases the asymptotic separation $\tilde L(\infty)$ of the D7--brane embeddings, and there exists an embedding which asymptotes to zero separation at infinity but has finite separation in the bulk of the geometry. Thus, the rotational symmetry along $\phi_3$ present at infinity ($\tilde L(\infty)=0$) is broken in the bulk, which corresponds to a spontaneous breaking of the dual $U(1)$ $R$--symmetry \cite{Evans:2004ia}. 
By generating the spectrum of fluctuations for various values of the bare mass parameter $\tilde m$ we will show that indeed the ground state is massless at vanishing bare mass. Furthermore, for small values of $\tilde m$ we will demonstrate a characteristic Gell-Mann-Oaks-Renner $\sqrt m_q $ dependence \cite{GellMann:1968rz} of the spectrum.

In the analysis, we have to take into account the non-trivial mixing of the $\Phi$-mode with the gauge field components $A_0, A_3$, or equivalently $F_{03}$, cf. (\ref{eomPhi}).
Again, we will study the meson spectrum numerically and require normalizability of the solutions in the UV. This condition will again lead to a discrete spectrum.

\begin{figure}[t] 
   \centering
   \includegraphics[width=5.0in]{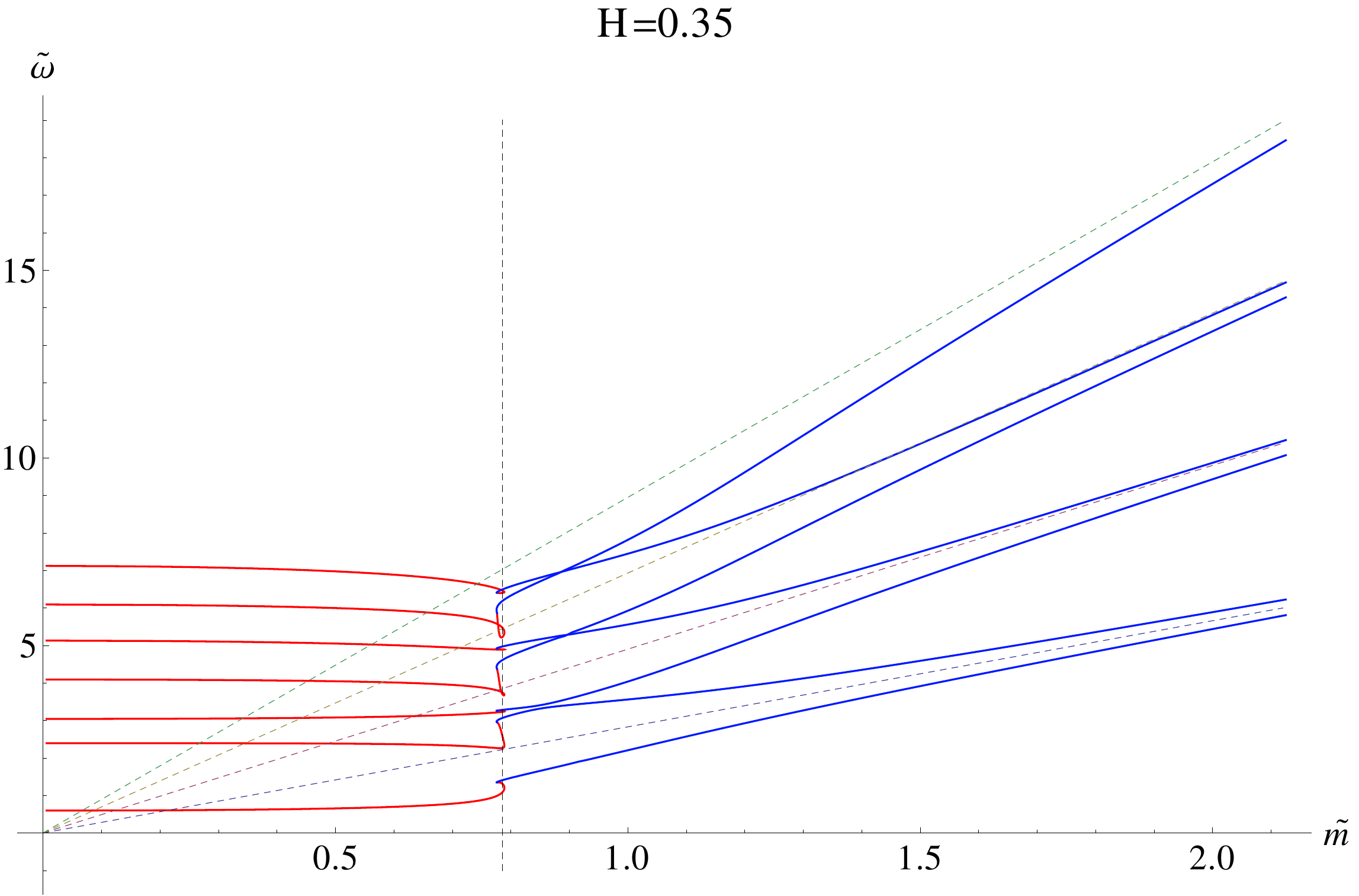} 
 \caption{The spectrum of $\Phi$ fluctuations vs. quark mass $\tilde{m}$ for various values of $H$. The vertical dashed line indicates the location of the critical bare mass
       $\tilde{m}_{\ast}$, while the coloured dashed lines correspond to the spectrum of pure $AdS_5 \times S^5$. The red curves correspond to the ball embeddings and 
        the blue curves represent the Minkowski embeddings; the phase transition happens for the critical embedding separating the two classes.}
   \label{fig:SpectrumPhi}
\end{figure}

\begin{figure}
   \centering
     \includegraphics[width=3.0in]{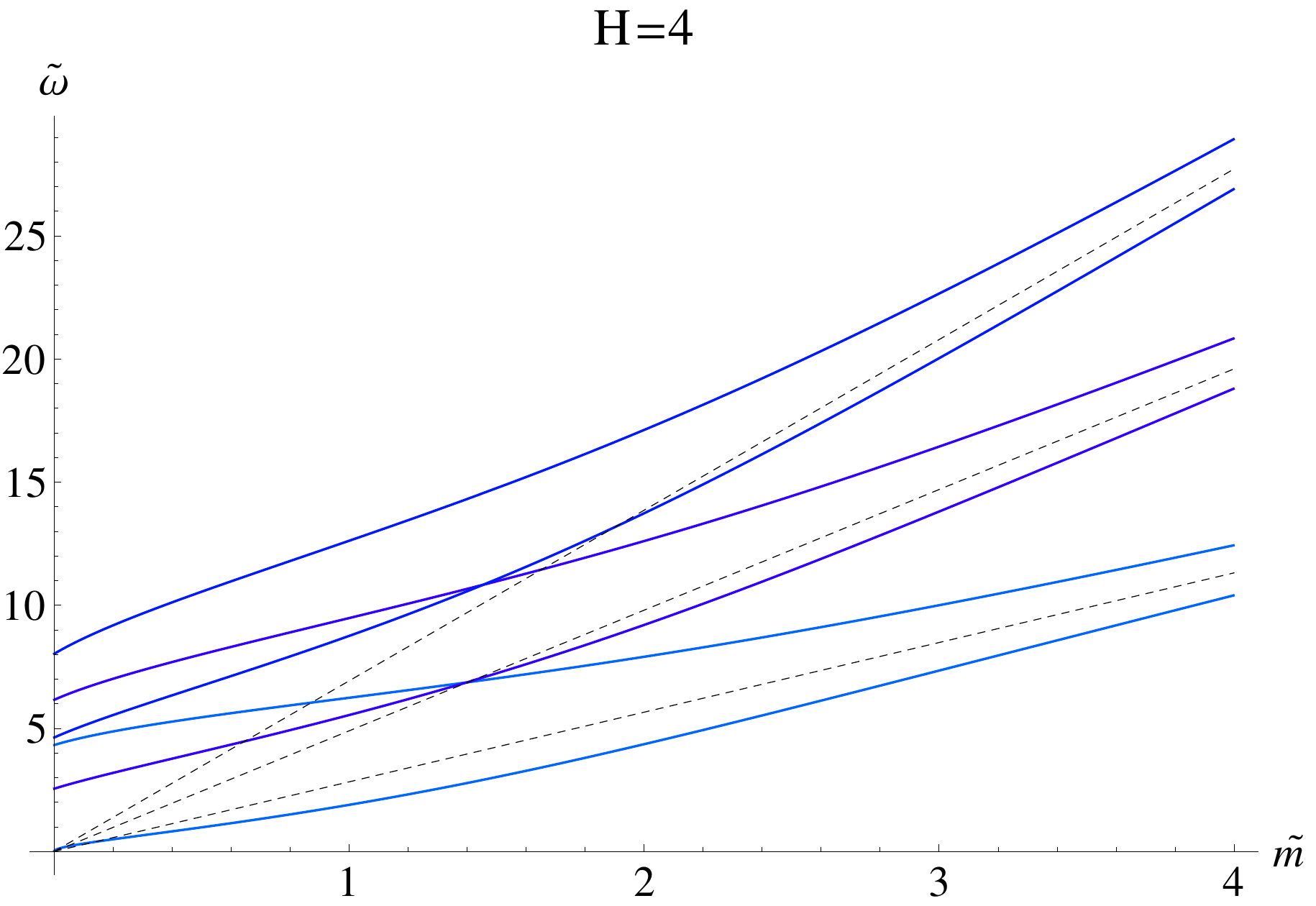} 
   \includegraphics[width=3.0in]{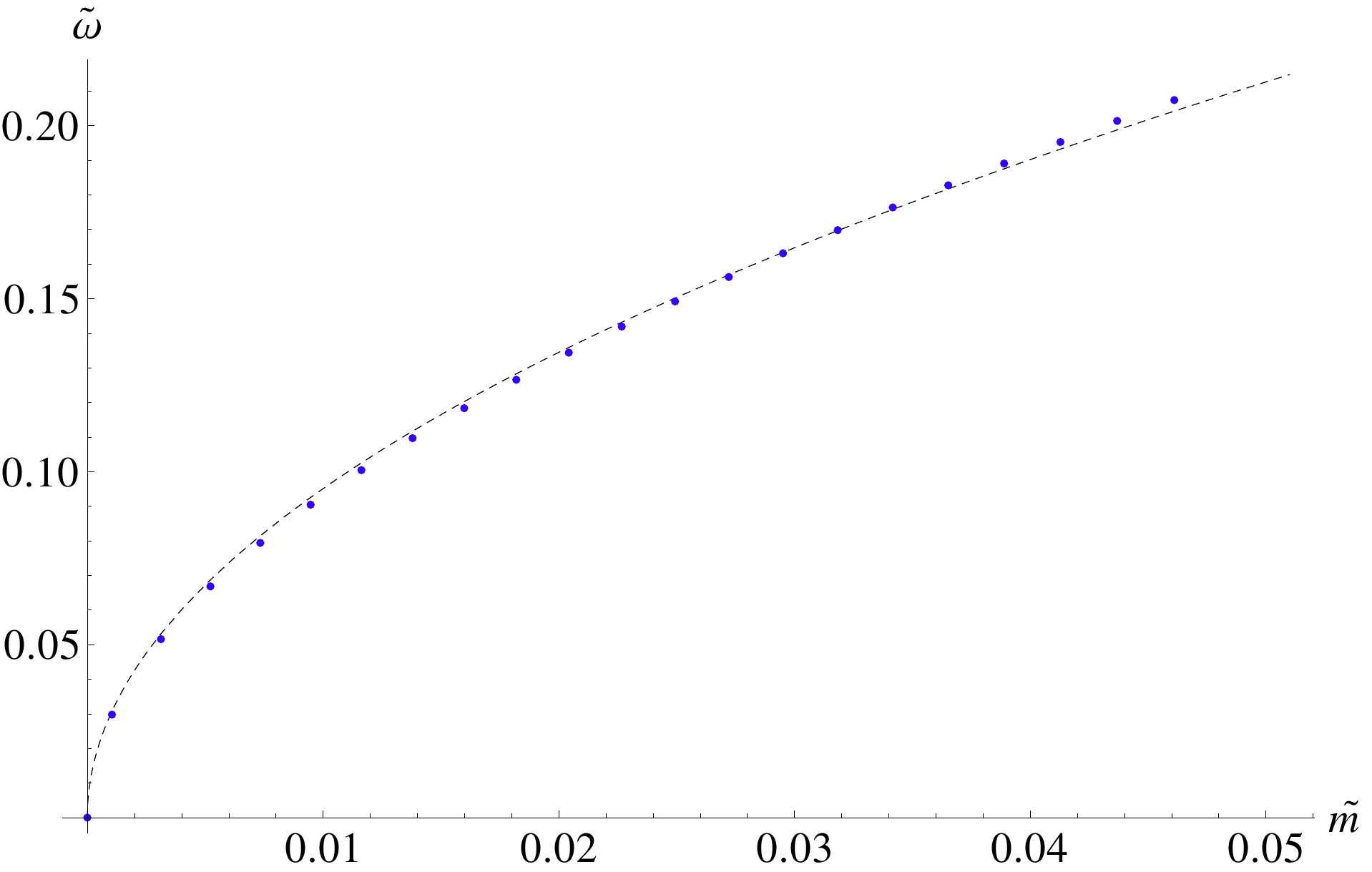} 
 \caption{The plot of the lowest lying $\Phi$ meson state for small bare mass exhibits the GMOR behaviour characteristic of a Goldstone boson.}
   \label{fig:GMORrel}
\end{figure}

The numerical results are presented in figure \ref{fig:SpectrumPhi}, where we plot $\tilde{\omega}$ versus $\tilde{m}$. For large bare quark mass $\tilde{m}$, the spectrum is expected to match the spectrum of the flat D3/D7 intersection on $AdS_5 \times S^5$, given by equation (\ref{spectflatcase}), and our results in the asymptotic region are indeed in good agreement with that expectation. 
As before, the red and blue curves correspond to the ``ball'' and ``Minkowski'' embeddings, respectively. One can see that near the phase transition (represented by the vertical dashed line in figure \ref{fig:SpectrumPhi}), the spectrum is multivalued and at the phase transition it has a finite jump. However, unlike the spectrum of fluctuations along $L$, there are no tachyonic modes in the self-similar region near the critical state.
As expected, our investigation confirms a Zeeman-like effect, namely a splitting of states (which will be proportional to the magnitude of $H$ at weak magnetic field or 
large $\tilde{m}$). This can be gleaned from figure \ref{fig:SpectrumPhi} where we observe two separate lines emanating from each asymptotic meson state at large 
bare mass $\tilde{m}$. At smaller values of $\tilde m$ the Zeeman splitting is strong and the energy levels intersect. This phenomenon is known as ``level crossing''.\\ \\
\noindent
{\bf Gell-Mann-Oakes-Renner (GMOR) relation.} In figure \ref{fig:GMORrel} we have presented the spectrum of fluctuations for strong magnetic field ($H=4.0>H_*$) when the phase transition disappears and theory is always in a confined phase. 
The spectrum is again featuring Zeeman splitting and intersection of energy levels (level crossing).
At vanishing bare mass the spectrum has a massless Goldstone mode. Zooming into the small bare mass region, the bare mass dependence of the ground state meson mass (the right plot in figure \ref{fig:GMORrel}) shows the characteristic Gell-Mann-Oaks-Renner relation,
$\tilde{M} \propto \sqrt{\tilde{m}}$, cf.~\cite{Filev:2009xp}. 

\section{Conclusions}
In this paper we studied the influence of an external magnetic field on a flavoured large $N_c$ gauge theory on $S^3$ in a semi-bottom up approach. 
We find that there is a competition between the effect of dissociation of the meson states due to the finite volume Casimir energy of the theory and 
the effect of the magnetic field which favours bound-states of mesons. 
As a result, the theory has an interesting phase diagram consisting of confined and deconfined phases separated by a critical curve across which there is a 
first order confinement/deconfinement quantum phase transition. At vanishing magnetic field the critical curve ends in a point where the phase transition 
is of third order\cite{Karch:2009ph}. 
For sufficiently strong magnetic fields, the phase transition seizes to exist and the theory is in the confined phase. 
In this regime the vacuum spontaneously breaks the global $U(1)$ $R$-symmetry by having a non-zero negative fundamental condensate. 
This is an example of magnetic catalysis of chiral symmetry breaking of a $U(1)$ axial symmetry. Thus,
the associated Goldstone boson is the analogue of $\eta'$ in QCD. The effect of the external magnetic field on the meson spectra is to couple the scalar and 
vector modes. At large bare masses the spectrum exhibit Zeeman splitting of the energy levels, which leads to level crossing in the confined phase of the theory. 
Across the phase transition the spectrum has a finite jump between the confined and deconfined phases. 
For sufficiently strong magnetic fields the only stable phase is the confined phase and the ground state of the spectrum possesses a massless mode 
corresponding to the Goldstone boson associated to the broken global $U(1)$ $R$-symmetry of the theory, in analogy to the $\eta'$ meson in QCD.
Furthermore by studying the dependence of the mass of the meson on the bare quark mass near the origin we have found a characteristic $M\propto\sqrt{m}$ Gell-Mann-Oakes-Renner relation.
A possible extension of our results would be to study the scenario with magnetic field at finite temperature. However, at finite temperature the phase
transition is thermal and the finite volume does not change the qualitative behaviour of the theory. Thus we expect the theory to have similar qualitative behaviour 
as in flat space. \\ 
An interesting direction for future work is to consider the effects of a magnetic field and various chemical potentials on fields theories with and without defects  
on compact manifolds. Defect field theories can be realized in an holographic framework, e.g., by the introduction of D5--brane probes \cite{Karch:2009ph,Erdmenger:2010zm}.

\section{Acknowledgements} 
The work of V.F. is funded by an INSPIRE IRCSET-Marie Curie International Mobility Fellowship and the work of M.I. is supported by an IRC EMPOWER Postdoctoral Fellowship. 
The authors are grateful to Tameem Albash, Andy O'Bannon and Stefano Kovacs for useful comments and discussions and would like to thank Johanna Erdmenger for collaboration 
in the early stages of this project.


\begin{thebibliography}{99}

\bibitem{Gusynin:1994re}
  V.~P.~Gusynin, V.~A.~Miransky and I.~A.~Shovkovy,
  Phys.\ Rev.\ Lett.\  {\bf 73}, 3499 (1994)
  [Erratum-ibid.\  {\bf 76}, 1005 (1996)]
  [arXiv:hep-ph/9405262].

\bibitem{Gusynin:1994xp}
  V.~P.~Gusynin, V.~A.~Miransky and I.~A.~Shovkovy,
  Phys.\ Lett.\  B {\bf 349}, 477 (1995)
  [arXiv:hep-ph/9412257].


\bibitem{Hong:1996pv}
  D.~K.~Hong, Y.~Kim and S.~J.~Sin,
  Phys.\ Rev.\  D {\bf 54}, 7879 (1996)
  [arXiv:hep-th/9603157].


\bibitem{Klimenko:1990rh}
  K.~G.~Klimenko,
  Theor.\ Math.\ Phys.\  {\bf 89}, 1161 (1992)
  [Teor.\ Mat.\ Fiz.\  {\bf 89}, 211 (1991)].


\bibitem{Klimenko:1991he}
  K.~G.~Klimenko,
  Z.\ Phys.\  C {\bf 54}, 323 (1992).


\bibitem{Klimenko:1992ch}
  K.~G.~Klimenko,
  Theor.\ Math.\ Phys.\  {\bf 90}, 1 (1992)
  [Teor.\ Mat.\ Fiz.\  {\bf 90}, 3 (1992)].


\bibitem{Maldacena:1997re}
  J.~M.~Maldacena,
  Adv.\ Theor.\ Math.\ Phys.\  {\bf 2}, 231 (1998)
  [Int.\ J.\ Theor.\ Phys.\  {\bf 38}, 1113 (1999)]
  [arXiv:hep-th/9711200].
  
\bibitem{Karch:2002sh} 
  A.~Karch and E.~Katz,
  JHEP {\bf 0206}, 043 (2002)
  [hep-th/0205236].

\bibitem{Filev:2007gb} 
  V.~G.~Filev, C.~V.~Johnson, R.~C.~Rashkov and K.~S.~Viswanathan,
  JHEP {\bf 0710}, 019 (2007)
  [hep-th/0701001].


\bibitem{Filev:2011mt} 
  V.~G.~Filev and D.~Zoakos,
  JHEP {\bf 1108}, 022 (2011)
  [arXiv:1106.1330 [hep-th]].
  
\bibitem{Erdmenger:2011bw} 
  J.~Erdmenger, V.~G.~Filev and D.~Zoakos,
  JHEP {\bf 1208}, 004 (2012)
  [arXiv:1112.4807 [hep-th]].
  
\bibitem{Ammon:2012qs} 
  M.~Ammon, V.~Filev, J.~Tarrio and D.~Zoakos,
  JHEP {\bf 1209}, 039 (2012)
  [arXiv:1207.1047 [hep-th]].
  
  
\bibitem{Chunlen:2014zpa} 
  S.~Chunlen, K.~Peeters, P.~Vanichchapongjaroen and M.~Zamaklar,
  arXiv:1405.1996 [hep-th].



\bibitem{Karch:2006bv}
  A.~Karch and A.~O'Bannon,
  Phys.\ Rev.\  D {\bf 74}, 085033 (2006)
  [arXiv:hep-th/0605120].

\bibitem{Karch:2009ph}
  A.~Karch, A.~O'Bannon and L.~G.~Yaffe,
  JHEP {\bf 0909} (2009) 042
  [arXiv:0906.4959 [hep-th]].
 
\bibitem{Erdmenger:2010zm} 
  J.~Erdmenger and V.~Filev,
  JHEP {\bf 1101}, 119 (2011)
  [arXiv:1012.0496 [hep-th]].
  
  
    
\bibitem{PremKumar:2011ag} 
  S.~P.~Kumar,
  Phys.\ Rev.\ D {\bf 84}, 026003 (2011)
  [arXiv:1104.1405 [hep-th]].
  
\bibitem{Chunlen:2012zy} 
  S.~Chunlen, K.~Peeters, P.~Vanichchapongjaroen and M.~Zamaklar,
  arXiv:1210.6188 [hep-th].
  
\bibitem{GellMann:1968rz} 
 M.~Gell-Mann, R.~J.~Oakes and B.~Renner,
 Phys.\ Rev.\ �{\bf 175}, 2195 (1968).
      
     
\bibitem{Karch:2005ms}
  A.~Karch, A.~O'Bannon and K.~Skenderis,
  JHEP {\bf 0604}, 015 (2006)
  [arXiv:hep-th/0512125].
  
    \bibitem{Filev:2008xt}
  V.~G.~Filev and C.~V.~Johnson,
  JHEP {\bf 0810}, 058 (2008)
  [arXiv:0805.1950 [hep-th]].
  
  \bibitem{Mateos:2007vn}
  D.~Mateos, R.~C.~Myers and R.~M.~Thomson,
  JHEP {\bf 0705}, 067 (2007)
  [arXiv:hep-th/0701132].
  
\bibitem{Mateos:2006nu}
  D.~Mateos, R.~C.~Myers and R.~M.~Thomson,
  Phys.\ Rev.\ Lett.\  {\bf 97}, 091601 (2006)
  [arXiv:hep-th/0605046].
  
  
\bibitem{Frolov:2006tc}
  V.~P.~Frolov,
  Phys.\ Rev.\  D {\bf 74}, 044006 (2006)
  [arXiv:gr-qc/0604114].
  
  
\bibitem{Albash:2007bk}
  T.~Albash, V.~G.~Filev, C.~V.~Johnson and A.~Kundu,
  JHEP {\bf 0807}, 080 (2008)
  [arXiv:0709.1547 [hep-th]].
    
\bibitem{Erdmenger:2007bn} 
  J.~Erdmenger, R.~Meyer and J.~P.~Shock,
  JHEP {\bf 0712}, 091 (2007)
  [arXiv:0709.1551 [hep-th]].
  
    
\bibitem{Filev:2009xp}
  V.~G.~Filev, C.~V.~Johnson and J.~P.~Shock,
  JHEP {\bf 0908}, 013 (2009)
  [arXiv:0903.5345 [hep-th]].
  
\bibitem{Arean:2005ar} 
  D.~Arean, A.~Paredes and A.~V.~Ramallo,
  JHEP {\bf 0508}, 017 (2005)
  [hep-th/0505181].
  
 
\bibitem{Kruczenski:2003be}
M.~Kruczenski, D.~Mateos, R.~C. Myers, and D.~J. Winters, 
   {\em JHEP} {\bf 07} (2003) 049,


\bibitem{Evans:2004ia} 
 N.~J.~Evans and J.~P.~Shock,
 Phys.\ Rev.\ D {\bf 70}, 046002 (2004)
 [hep-th/0403279].

  
\bibitem{Myers:2006qr}
  R.~C.~Myers and R.~M.~Thomson,
  JHEP {\bf 0609}, 066 (2006)
  [arXiv:hep-th/0605017].

\bibitem{Filev:2007qu} 
  V.~G.~Filev,
  JHEP {\bf 0804}, 088 (2008)
  [arXiv:0706.3811 [hep-th]].
  
\bibitem{Erdmenger:2007cm}
  J.~Erdmenger, N.~Evans, I.~Kirsch and E.~Threlfall,
  Eur.\ Phys.\ J.\  A {\bf 35}, 81 (2008)
  [arXiv:0711.4467 [hep-th]].
 
\bibitem{Johnson:2008vna}
  C.~V.~Johnson and A.~Kundu,
  JHEP {\bf 0812}, 053 (2008)
  [arXiv:0803.0038 [hep-th]].
  
\bibitem{Bergman:2008sg} 
 O.~Bergman, G.~Lifschytz and M.~Lippert,
 JHEP {\bf 0805}, 007 (2008)
 [arXiv:0802.3720 [hep-th]].
 
  
\bibitem{Bergman:2008qv} 
 O.~Bergman, G.~Lifschytz and M.~Lippert,
 Phys.\ Rev.\ D {\bf 79}, 105024 (2009)
 [arXiv:0806.0366 [hep-th]].
  
\bibitem{Zayakin:2008cy}
  A.~V.~Zayakin,
  JHEP {\bf 0807} (2008) 116
  [arXiv:0807.2917 [hep-th]].
   
\bibitem{Filev:2009ai}
  V.~G.~Filev,
  JHEP {\bf 0911}, 123 (2009)
  [arXiv:0910.0554 [hep-th]].

\bibitem{Evans:2010iy}
  N.~Evans, A.~Gebauer, K.~Y.~Kim and M.~Magou,
  JHEP {\bf 1003}, 132 (2010)
  [arXiv:1002.1885 [hep-th]].
  
  
\bibitem{Jensen:2010vd}
  K.~Jensen, A.~Karch and E.~G.~Thompson,
  JHEP {\bf 1005}, 015 (2010)
  [arXiv:1002.2447 [hep-th]].
  
\bibitem{Jensen:2010ga}
  K.~Jensen, A.~Karch, D.~T.~Son and E.~G.~Thompson,
  Phys.\ Rev.\ Lett.\  {\bf 105}, 041601 (2010)
  [arXiv:1002.3159 [hep-th]].
  
\bibitem{Evans:2010hi}
  N.~Evans, A.~Gebauer, K.~Y.~Kim and M.~Magou,
  Phys.\ Lett.\  B {\bf 698}, 91 (2011)
  [arXiv:1003.2694 [hep-th]].
  
\bibitem{Filev:2010pm}
  V.~G.~Filev and R.~C.~Rashkov,
  Adv.\ High Energy Phys.\  {\bf 2010}, 473206 (2010)
  [arXiv:1010.0444 [hep-th]].
  
  
\bibitem{Bayona:2010bg} 
  C.~A.~B.~Bayona, H.~Boschi-Filho, M.~Ihl and M.~A.~C.~Torres,
  JHEP {\bf 1008}, 122 (2010)
  [arXiv:1006.2363 [hep-th]].
\bibitem{Ihl:2010zg} 
 M.~Ihl, M.~A.~C.~Torres, H.~Boschi-Filho and C.~A.~B.~Bayona,
 JHEP {\bf 1109}, 026 (2011)
 [arXiv:1010.0993 [hep-th]].
 

\bibitem{Evans:2011mu} 
  N.~Evans, A.~Gebauer and K.~Y.~Kim,
  JHEP {\bf 1105}, 067 (2011)  [arXiv:1103.5627 [hep-th]].  
  
\bibitem{Callebaut:2011ab} 
  N.~Callebaut, D.~Dudal and H.~Verschelde,
  arXiv:1105.2217 [hep-th].
  

\bibitem{Bolognesi:2011un} 
  S.~Bolognesi and D.~Tong,
  arXiv:1110.5902 [hep-th].  
  
\bibitem{Bolognesi:2012pi} 
 S.~Bolognesi, J.~N.~Laia, D.~Tong and K.~Wong,
 JHEP {\bf 1207}, 162 (2012)
 [arXiv:1204.6029 [hep-th]].
 
 \bibitem{Ihl:2012bm} 
  M.~Ihl, A.~Kundu and S.~Kundu,
  JHEP {\bf 1212}, 070 (2012)
  [arXiv:1208.2663 [hep-th]].
  
 \bibitem{Alam:2013cia} 
  M.~Sohaib Alam, M.~Ihl, A.~Kundu and S.~Kundu,
  JHEP {\bf 1309}, 130 (2013)
  [arXiv:1306.2178 [hep-th]].
 
 \bibitem{Bali:2011qj} 
  G.~S.~Bali, F.~Bruckmann, G.~Endrodi, Z.~Fodor, S.~D.~Katz, S.~Krieg, A.~Schafer and K.~K.~Szabo,
  JHEP {\bf 1202}, 044 (2012)
  [arXiv:1111.4956 [hep-lat]].
 \bibitem{Herzog:2010uh}
  C.~P.~Herzog, S.~A.~Stricker and A.~Vuorinen,
  Phys.\ Rev.\ D {\bf 82} (2010) 041701
  [arXiv:1005.3285 [hep-th]].
\bibitem{Chernodub:2010qx}
  M.~N.~Chernodub,
  Phys.\ Rev.\ D {\bf 82} (2010) 085011
  [arXiv:1008.1055 [hep-ph]],
  M.~N.~Chernodub,
  Phys.\ Rev.\ Lett.\  {\bf 106}, 142003 (2011)
  [arXiv:1101.0117 [hep-ph]],
  V.~V.~Braguta, P.~V.~Buividovich, M.~N.~Chernodub and M.~I.~Polikarpov,
  arXiv:1104.3767 [hep-lat].
  

  
  



  

  

  
\end{thebibliography}
\end{document}